\documentclass[twocolumn,superscriptaddress,amssymb,amsmath,nobibnotes,aps,prd,showpacs,showkeys,nofootinbib]{revtex4-1}
\pdfoutput=1

\usepackage{graphicx,subfigure,bm,color,psfrag,hyperref}
\usepackage{amsfonts}
\usepackage{lipsum}
\usepackage{caption}
\usepackage{mathtools}
\usepackage{verbatim}
\usepackage[bottom]{footmisc}
\usepackage{xcolor}

\newcommand{\be}{\begin{equation}}
\newcommand{\ee}{\end{equation}}
\usepackage{changes}
\begin{document}

\title{Forecast constraints on $f(T)$ gravity with gravitational waves from compact binary coalescences}

\author{Rafael C. Nunes}
\email{rafadcnunes@gmail.com}
\affiliation{Divis\~ao de Astrof\'isica, Instituto Nacional de Pesquisas Espaciais, Avenida dos Astronautas 1758, S\~ao Jos\'e dos Campos, 12227-010, SP, Brazil}

\author{M\'arcio E. S. Alves}
\email{marcio.alves@unesp.br}
\affiliation{Universidade Estadual Paulista (UNESP), Instituto de Ci\^encia e Tecnologia \\ S\~ao Jos\'e dos Campos, SP, 12247-004, Brazil}

\author{Jos\'e C. N. de Araujo}
\email{jcarlos.dearaujo@inpe.br}
\affiliation{Divis\~ao de Astrof\'isica, Instituto Nacional de Pesquisas Espaciais, Avenida dos Astronautas 1758, S\~ao Jos\'e dos Campos, 12227-010, SP, Brazil}

\begin{abstract}

The direct detection of gravitational waves (GWs) opened a new chapter in the modern cosmology to probe possible deviations from the general relativity (GR) theory. In the present work, we investigate for the first time the modified GW form propagation from the inspiraling of compact binary systems within the context of $f(T)$ gravity in order to obtain new forecasts/constraints on the free parameter of the theory. First, we show that the modified waveform differs from the GR waveform essentially due to induced corrections on the GWs amplitude. Then, we discuss the forecasts on the $f(T)$ gravity assuming simulated sources of GWs as black hole binaries, neutron star binaries and black hole - neutron star binary systems, which emit GWs in the frequency band of the Advanced LIGO (aLIGO) interferometer and of the third generation Einstein Telescope (ET). We show that GWs sources detected within the aLIGO  sensitivity can return estimates of the same order of magnitude of the current cosmological observations. On the other hand, detection within the ET sensitivity can improve by up to 2 orders of magnitude the current bound on the $f(T)$ gravity. Therefore, the statistical accuracy that can be achieved by future ground based GW observations, mainly with the ET detector (and planed detectors with a similar sensitivity), can allow strong bounds on the free parameter of the theory, and can be decisive to test the theory of gravitation.
\end{abstract}


%---------------------------------------------------
\maketitle
%---------------------------------------------------
\section{Introduction}

The detection of gravitational waves (GWs) in recent years opened a new spectrum of possibilities to investigate cosmic phenomena and fundamental physics (see \cite{LIGO01} for a summary of all GWs detection up to the present time). Beyond the present performance of the LIGO and Virgo interferometers, with a third generation of detectors the precision GW astronomy will become a reality. As remarkable examples, we mention a planned third generation of the LIGO interferometer (called LIGO Voyager \cite{LVoyager}) and the Einstein Telescope (ET) \cite{ET_design01,ET_design02}. Such GW observatories will allow for observations with signal-to-noise ratios that are several times larger than the current aLIGO.

Certainly, current and future GW observations will play important role in unraveling open problems in modern cosmology, as a complementary source of information to probe dark energy and modified theories of gravity. After the recent GW170817 and GRB 170817A events, the modified gravity theories have been screened by imposing strong constraints over them \cite{GW_MG01,GW_MG02,GW_MG03,GW_MG04,GW_MG041} (see also \cite{GW_MG05} for the latest review), following an exclusion principle applicable to those models which predict that the speed of GWs is not equal to the speed of light.

Furthermore, the detection of GWs from the merger of compact binary systems, as black holes and neutron star binaries, allows one to infer the luminosity distance without the need of a calibration with respect to another source. In this sense, these systems have been called ``standard sirens''. Therefore, if the redshift of the GW sources are measured by using other techniques, the GW observations will provide an independent tool to constraint the expansion history of the Universe and to test alternative theories of gravity in a cosmological setting.

Recently, it has been shown that modifications in the underlying gravity theory can affect not only the speed and the  waveform of the GWs, but also their propagation through cosmological distances. Hence, in the context of modified theories of gravity a new equation for the GW luminosity distance arises, which is distinct from its electromagnetic version \cite{MG_amplitude01}. Therefore, such an effect is a new opportunity to test or even to rule out alternative theories of gravity \cite{MG_amplitude01, MG_amplitude02}. In the present article, such corresponding corrections on $f(T)$ gravity are presented for the first time.

Amongst various modified gravity theories available at the present \cite{MG01,MG02}, the $f(T)$-teleparallel modified gravity theories gained a massive interest in the scientific community \cite{fT01, fT02,fT03,fT04,fT05,fT06,fT07,fT08,fT09,fT10,fT11,fT12,fT13,fT14,fT15,fT16,fT17,fT18,fT19,fT20,fT21,fT22, fT23, fT24,fT25,fT26,fT27,fT28} due to its dual ability to mimick the early and late accelerating phases of the Universe without the inclusion of a dark energy fluid (see \cite{Cai:2015emx} for review). The $f(T)$ theory is a class of torsion-based modified gravity theories where the torsion scalar, $T$, plays an equivalent role to the Ricci scalar in the Einstein-Hilbert action.

Recent theoretical developments in $f(T)$ gravity in the context of GWs were made in \cite{Cai:2018rzd,Farrugia:2018gyz, HohmannGW1, HohmannGW2}. In particular, the first extraction of the observational constrains using GWs physics through a stochastic primordial GWs within the context of $f(T)$-teleparallel modified gravity was presented in \cite{Eu}. On the other hand, connections between observation and theory in the context of GWs from compact binaries within the $f(T)$ gravity/cosmology have not been investigated so far, and here we present such results for the first time in the literature, which we believe to be one of the most important issues to analyze when treating phenomenological scenarios and their feasibility.
In the present work, we obtain forecasts on the parameter estimation in the $f(T)$ gravity for observations of compact binary coalescences by the Advanced LIGO (aLIGO) and for ET. In general, as we will see later the main effect of the $f(T)$ gravity is to modify the GW propagation between the GWs source and detector. Therefore, we focus on the modified GW luminosity distance obtained in the context of the $f(T)$ gravity, which is a cumulative effect over the distance to the source \cite{Yunes2016}.

The manuscript is organized as follows:  In the next section, we present the equations that determine the propagation of GWs in the $f(T)$ gravity. In Section \ref{Modified_propagation}, we quantify the modification in the GW form compared with the GR theory. In Section \ref{Fisher}, we summarize the statistical methodology used in the parameter estimation. In Section \ref{results} we present the forecasts for GW observations with the LIGO em ET detectors. Finally,  we present our concluding remarks and perspectives in Section \ref{Conclusions}.

\section{Tensor perturbation in $f(T)$ gravity}
\label{sec-model}

In this section we briefly review the main aspects of the $f(T)$ gravity, the resulting Friedmann
equations as well as the equations of motion of the tensor modes.

In the framework of torsional gravity one uses the tetrad fields
$e^\mu_A$ as the dynamical variables, which form an orthonormal base in the tangent space of the underlying manifold $\mathcal{M}$, which is endowed with a metric tensor $g_{\mu\nu}=\eta_{A B} e^A_\mu e^B_\nu$, where the tetrad metric is $\eta_{A B} = {\rm diag}(-1, 1, 1, 1)$. Throughout this work we use Greek indices to denote 
the coordinate space and Latin capital indices for the tangent space. Furthermore, 
unlike the GR theory, which uses the torsionless Levi-Civita connection, here we use the curvatureless Weitzenb{\"{o}}ck connection $\overset{\mathbf{w}}{\Gamma}^\lambda_{\nu\mu}\equiv
e^\lambda_A\:\partial_\mu e^A_\nu$  \cite{Pereira.book}. Thus, 
the gravitational field is described by the torsion tensor
\begin{equation}\label{energy-momentum}
T^\lambda_{\verb| |\mu\nu} \equiv \overset{\mathbf{w}}{\Gamma}^\lambda_{\nu\mu} - \overset{\mathbf{w}}{\Gamma}^\lambda_{\mu\nu}  = e^\lambda_A
\left( \partial_\mu e^A_\nu - \partial_\nu e^A_\mu \right).
\end{equation} 

In the specific case of the teleparallel equivalent of general relativity (TEGR), the Lagrangian is the torsion scalar $T$, constructed as follows \cite{Pereira.book}
\begin{equation}
\label{T-scalar}
T\equiv\frac{1}{4}
T^{\rho \mu \nu}
T_{\rho \mu \nu}
+\frac{1}{2}T^{\rho \mu \nu }T_{\nu \mu\rho}
-T_{\rho \mu}{}^{\rho }T^{\nu\mu}{}_{\nu}\, \, ,
\end{equation}
and the corresponding  action reads
\begin{equation}
S= \frac{1}{16 \pi G} \int d^4x\, e\, T ,    
\end{equation}
where $e = \text{det}(e_{\mu}^A) = \sqrt{-g}$ and
$G$ is the Newton's gravitational constant (we set the speed of light to $c =1$). 

If we use TEGR as the starting point for torsional modified gravity, the simplest modification is the action for the $f(T)$ gravity, which is given by
\begin{align}
\label{action}
 S= \frac{1}{16 \pi G} \int d^4x\, e\, f(T) ~.
\end{align}

The variation of the above action with
respect to the tetrads leads to the field equations, namely 
\begin{align}
\label{field-eqs}
 & e^{-1}\partial_{\mu} (ee_A^{\rho}S_{\rho}{}^{\mu\nu}) f_{T} + e_A^{\rho}
S_{\rho}{}^{\mu\nu} \partial_{\mu}({T}) f_{TT} \nonumber\\
 & - f_{T} e_{A}^{\lambda} T^{\rho}{}_{\mu\lambda} S_{\rho}{}^{\nu\mu} + \frac{1}{4} e_
{A}^{\nu} f(
{T}) %\nonumber \\
 = 4 \pi G e_{A}^{\rho} \Theta_{\rho}{}^{\nu} ~,
\end{align}
where $f_{T}=\partial f/\partial T$, $f_{TT}=\partial^{2} f/\partial T^{2}$, and 
$\Theta_{\rho}
{}^{\nu}$  denotes the energy-momentum tensor of the matter sector. In Eq. (\ref{field-eqs}) 
we have introduced, for convenience, the
``super-potential'', namely  
\begin{equation}
S_\rho^{\:\:\:\mu\nu} \equiv \frac{1}{2} \left( {\cal{K}}^{\mu\nu}_{\:\:\:\:\:\rho} +
\delta^\mu_\rho \: T^{\alpha\nu}_{\:\:\:\:\:\alpha} - \delta^\nu_\rho \:
T^{\alpha\mu}_{\:\:\:\:\:\:\alpha} \right),
\end{equation}
where the contorsion tensor is defined as follows 
\begin{equation}
\label{contor}
{\cal{K}}_{\:\:\mu\nu}^{\rho} \equiv \frac{1}{2} \left( T_{\mu\: \: \: \nu}^{\:\:\:\rho} + T_{\nu\: \: \: \mu}^{\:\:\:\rho} - T_{\:\:\mu\nu}^{\rho} \right).
\end{equation}

Applying the $f(T)$ gravity in a cosmological framework and imposing the homogeneous 
and isotropic geometry, we have the diagonal tetrad $e_{\mu}^A={\text{diag}}\left(1,a(t),a(t),a(t) \right)$. 
This vierbein corresponds to the spatially flat Friedmann-Lema\^itre-Robertson-Walker (FLRW)  metric, namely
\begin{equation}
ds^2= dt^2-a^2(t)\,  \delta_{ij} dx^i dx^j,
\end{equation}
with $a(t)$ the scale factor; and inserting it into the general field equations 
(\ref{field-eqs}), we obtain the following Friedmann equations 
\begin{eqnarray}\label{Fr11}
&&H^2= \frac{8\pi G}{3}\rho_m
-\frac{f}{6}+\frac{Tf_T}{3},\\\label{Fr22}
&&\dot{H}=-\frac{4\pi G(\rho_m+p_m)}{1+f_{T}+2Tf_{TT}}.
\end{eqnarray}

In the above equations $H\equiv\dot{a}/a$ is the Hubble function, with dots denoting
derivatives with respect to $t$, and $\rho_m$ and $p_m$ are the energy density 
and pressure for the matter perfect fluid, respectively.

Let us now study the perturbations of the $f(T)$ gravity around the FLRW cosmological background, focusing on the tensor sector. 

Following \cite{Cai:2018rzd} the tetrad can be decomposed as $e^A_{\mu}(x) = \bar{e}^A_{\mu}(x) + \chi^A_{\mu}(x)$, which satisfies the equation $g_{\mu\nu}(x) = \eta_{AB} e^A_{\mu} e^B_{\nu} = \eta_{AB} \bar{e}^A_{\mu} \bar{e}^B_{\nu}$, where  $\bar{e}^A_{\mu}$ represents the part of the tetrad corresponding to metric components, and $\chi^A_{\mu}$ represents the degrees of freedom related to local Lorentz transformation. Therefore, in what follows we focus only on $\bar{e}^A_{\mu}(x)$. Actually, the full version of the teleparallel gravity and its modification need to include the spin connection (see \cite{Krssak}). Then, for a $f(T)$ theory with tetrads and spin
connections, we vary the action with respect to the tetrads and obtain a set of equations about the tetrads and spin connection. This set of equations is local Lorentz invariance since the change of tetrads given by Lorentz transformation is "eaten" by the spin connection. In order to solve these equations in a easy way, we choose a frame that let spin connection is zero. This means we break the local Lorentz invariance by our own hand. Thus, the additional components of tetrads such as $\chi^A_\mu$
are present. See also the recent work \cite{fT16} for a more extended discussion on this.

Now, the tetrad $\bar{e}^A_{\mu}(x)$ can be perturbed around the FLRW geometry in scalar, vector and tensor modes. Since we are interested in GWs, let us consider only the transverse traceless tensor mode $h_{ij}$ and set the other modes to zero. In this case, the perturbed tetrad reads 
\begin{align}
\label{perturbation}
 \bar{e}^0_{\mu} =& \delta^0_{\mu}~,
\nonumber \\
 \bar{e}^a_{\mu} =& a \left( \delta^a_{\mu} + \frac{1}{2} \delta^i_{\mu}  \delta^{aj}   h_{ij} \right) ~, \nonumber \\
 \bar{e}_0^{\mu} =& \delta_0^{\mu} ~, \nonumber \\
 \bar{e}_a^{\mu} =& \frac{1}{a} \left( \delta_a^{\mu}  - \frac{1}{2} \delta^{\mu i} \delta^j_a   h_{ij} \right) ~,
\end{align}
which leads to the usual perturbed metric, namely
\begin{align}
g_{00} = -1,~~~g_{i0} = 0,~~~g_{ij} = a^2(t)h_{ij}.    
\end{align}

Inserting (\ref{perturbation}) into (\ref{energy-momentum}) we obtain the components of the torsion tensor perturbed to first order
\begin{align}
 T^i{}_{0j} = & H\delta_{ij} + \frac12 \dot{h}_{ij} \nonumber \\
 T^i{}_{jk} = & \frac12 \left( \partial_j h_{ik} - \partial_k h_{ij} \right) ~,
\end{align}
and, on the other hand, the torsion scalar (\ref{T-scalar}) is unaffected at the linear order reading
\begin{align}
T =T^{(0)}+O(h^2),
\end{align}
with $T^{(0)} = 6H^2$ the zeroth-order torsion scalar. Such a result is valid even if the scalar and vector perturbations are included, this lies behind the fact that in the $f(T)$ gravity the GWs do not have extra polarization modes \cite{Farrugia:2018gyz}. Moreover,
the perturbed super-potential  can be written as 
\begin{equation}
S_i{}^{0j} = H\delta_{ij} -\frac14\dot{h}_{ij},~{\rm and}~~S_i{}^{jk} = \frac{1}{4a^2} (\partial_jh_{ik} -\partial_k h_{ij}).   
\end{equation}
 
Now, inserting the perturbed quantities into the field equations (\ref{field-eqs}), assuming that the background Friedmann equations (\ref{Fr11}) and (\ref{Fr22}) are satisfied and that there is no anisotropic stress contribution from the energy-momentum tensor, we obtain the equations of motion of GWs in the $f(T)$ cosmology, namely 
\begin{equation}
\label{gw_ft_0}
\ddot{h}_{ij} + 3H\left(1 - \beta_T\right)\dot{h}_{ij} - \frac{1}{a^2}\nabla^2 h_{ij} = 0,    
\end{equation} 
where the derivative $f_{T}$ is calculated at $T=T^{(0)}$, and we have introduced the dimensionless parameter  \cite{Cai:2018rzd}
\begin{equation}
\label{beta_T}
\beta_T \equiv - \frac{\dot{f}_T}{3 H f_T}.
\end{equation}

Finally, Eq. (\ref{gw_ft_0}) can be rewritten in terms of the conformal time\footnote{The corformal time $\tau$ is related to the cosmic time by $d\tau = dt /a(t)$.} as follows

\begin{equation}
\label{gw_ft}
\tilde{h}''_{ij} + 2  \mathcal{H}\left( 1 - \frac{3}{2} \beta_T\right) \tilde{h}'_{ij} + k^2 \tilde{h}_{ij} = 0,
\end{equation}
where we have assumed that the Fourier modes of the tensor perturbations $\tilde{h}_{ij}$ obey the Helmholtz equation $\nabla^2 \tilde{h}_{ij} + k^2 \tilde{h}_{ij} = 0$, with the wavenumber $k$. Therefore, as first showed in \cite{Cai:2018rzd}, from the above equation one can deduce that  the  speed of GWs is equal to the speed of light, and thus the experimental constraint of GW170817 is trivially satisfied in the $f(T)$ gravity.   

 \section{Modified GWs propagation in $f(T)$ gravity}
\label{Modified_propagation}

Since the GW amplitude is inversely proportional to the luminosity distance, the modification in the amplitude that comes from Eq. (\ref{gw_ft}) can be interpreted as a correction to the GW luminosity distance on general theories of modified gravity, as argued in Refs. \cite{MG_amplitude01, MG_amplitude02}. The effective luminosity distance, or equivalently an effective GW amplitude correction has been recently investigated in several contexts of modified gravity (see, e.g., Refs. \cite{Atsushi, MG_amplitude03, MG_amplitude04, MG_amplitude05, MG_amplitude06,MG_amplitude07,MG_amplitude08,MG_amplitude09,MG_amplitude10,MG_amplitude11,MG_amplitude12}). In this section, we are presenting the corresponding version, in the context of $f(T)$ gravity, for the first time.

Considering the formulation of the GW propagation and following the methodology presented in \cite{MG_amplitude01, MG_amplitude02}, a general expression for the effective luminosity distance in $f(T)$ gravity can be writen as
\begin{equation}
\label{dL_Gw_ft}
d_L^{GW} = d_L^{EM} \exp \Big[-\frac{3}{2}\int_0^{z} \frac{dz'}{1 + z'} \beta_T(z') \Big],
\end{equation}
where $\beta_T$ was previously defined in Eq. (\ref{beta_T}). Here, $d_L^{EM}$ is the standard luminosity distance for an electromagnetic signal, namely

\begin{equation}
\label{dL_em}
d_L^{EM}  = (1 + z) \int_0^z \frac{dz'}{H(z')},
\end{equation}
which is the same for gravitational radiation in the GR theory.

\begin{figure}
\includegraphics[width=3.5in,height=2.5in]{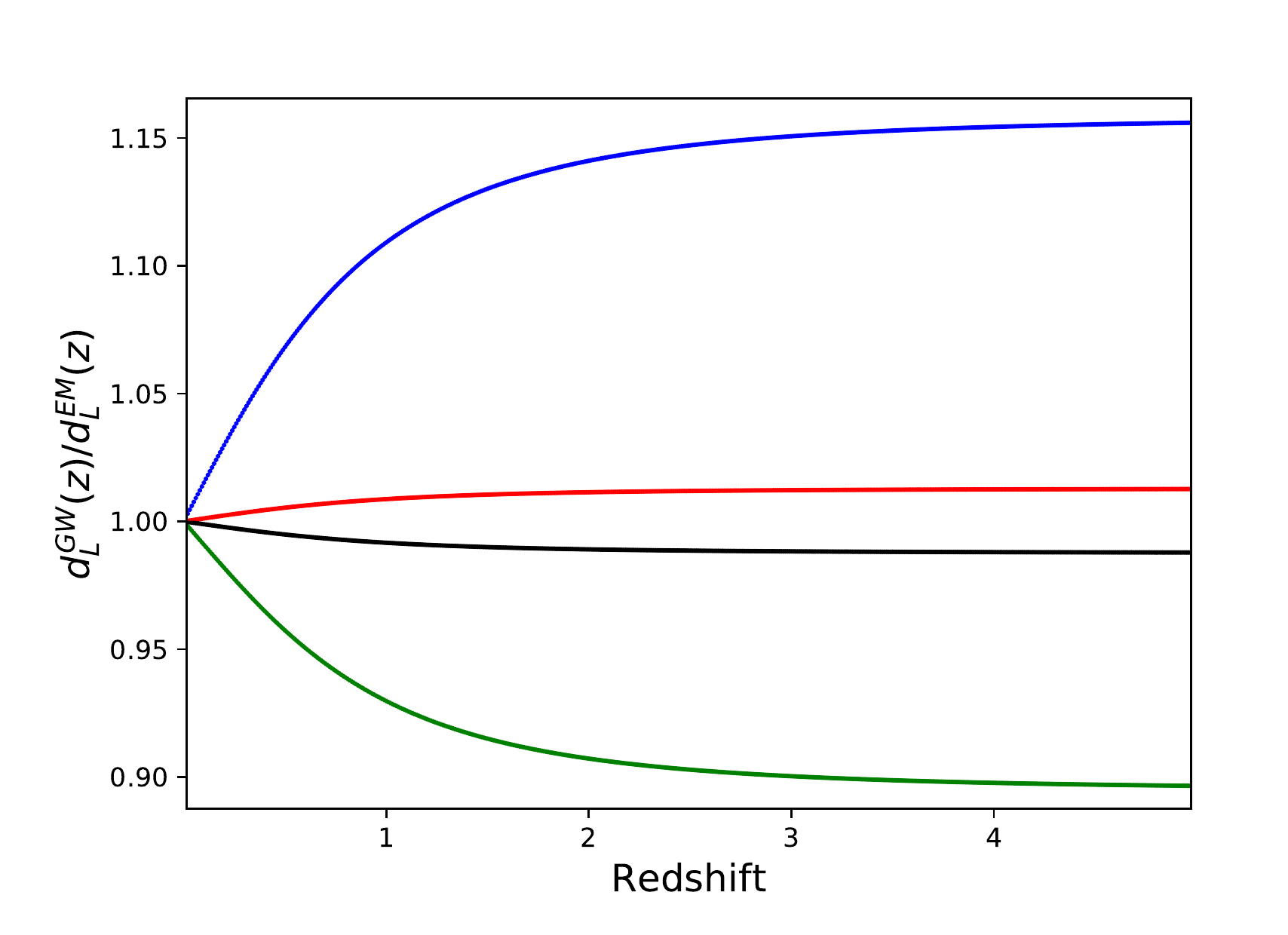}
\caption{The term $d_L^{GW}/d_L^{EM}$ quantifying the deviations with respect to GR is shown as a function of the redshift $z$ for the following selected values of $b = -0.01, -0.001, 0.001$, and 0.01 in green, black, red and blue, respectively.}
\label{delta_dL}
\end{figure}

In order to move on, we need to specific a model of $f(T)$ gravity. Without loss of generality, we will consider the power-law one, which is one of the most viable for cosmology, although our methodology can be applied for any $f(T)$ functional form. The power-law scenario corresponds to  \cite{fT04}:

\begin{equation}
\label{the_model}
f(T) = T + \alpha(-T)^b,
\end{equation}
where $\alpha$ and $b$ are parameters. Inserting (\ref{the_model})  into  (\ref{Fr11}) at present time we obtain
\begin{eqnarray}
\alpha=(6H_0^2)^{1-b}\left(\frac{1-\Omega_{m0}-\Omega_{r0}}{2b-1}\right),
\end{eqnarray}
with $\Omega_{m0}$ and $\Omega_{r0}$ are respectively the current values of the matter and 
radiation density parameters, and $H_0$ is the present value of the Hubble parameter. Thus, the only additional free parameter of the theory is $b$. By taking $b = 0$ one recovers GR (i.e., $\Lambda$CDM cosmology). 

Figure \ref{delta_dL} shows the GWs correction on the effective luminosity distance as a function of redshift $z$ for some selected values of $b$. One can notice that positive (negative) values of $b$ induce 
$d_L^{GW} > d_L^{EM}$ ($d_L^{GW} < d_L^{EM}$). Therefore, since the amplitude of GWs is $\propto 1/ d_L^{GW}$, $b  > 0$ $(< 0)$ causes a decrease (increase) in the GW amplitude. We found that within the range of values of $b$ we have chosen, the differences in $d_{L}$ are less than 15\% for all $z$. Notice also that for $z << 1$ we have $d_L^{GW} \simeq d_L^{EM}$, and hence significant deviations from GR are not expected for small redshifts.

\section{Parameter estimation}
\label{Fisher}

In what follows, we briefly introduce the foundations of a Fisher analysis and define the calculations used to find the estimated bounds on our free parameters, which will be presented in the next section.

The most accurate way to determine statistical limits on new free parameters (additional parameters with respect to a fiducial model, in our case, the GR theory) is through a Bayesian analysis. In such an approach, one calculates the full posterior probability distribution of the full parameter baseline of the model, given a set of observations/experiments. Within the GW context, for a high enough signal-to-noise ratio (SNR) \cite{Fisher01, Fisher02}, an approximation to the Bayesian procedure, a Fisher analysis, can be used to provide upper bounds for the free parameters of the models by means of the Cramer-Rao bound \cite{Fisher03, Fisher04}. Also, a Fisher analysis is useful to investigate forecasts on future experiments \cite{Fisher05}, as is the case of the present work.

In a Fisher analysis, one assumes that the likelihood probability
function has a single Gaussian peak, and approximates the behavior of the signal about that peak through a Taylor expansion. The result is a measure of the variance and of the covariance of the parameters in the template model through integrals that depend only on the templates and on the spectral noise density of the detector. In what follows, we summarize the main details of the calculation. We refer the reader to Refs. \cite{Fisher06, Fisher07, Fisher08, Fisher09, Fisher10} for a discussion and implementation of the Fisher analysis to estimate parameters in binary systems, which are the systems we consider in the present article.

Given a waveform model, $h(f, \theta_i )$, with the free parameters $\theta_i$, the root-mean-squared error on any parameter is determined by

\begin{align}
\label{}
\Delta \theta^i = \sqrt{\Sigma^{ii}},
\end{align}
where $\Sigma^{ij}$ is the covariance matrix, i.e, the inverse of the Fisher matrix, $\Sigma^{ij} = \Gamma_{ij}^{-1}$. The Fisher matrix is given by
\begin{align}
\label{}
\Gamma_{ij} = \Big( \frac{\partial \tilde{h}}{\partial \theta^i} \mid  \frac{\partial \tilde{h}}{\partial \theta^j}  \Big),
\end{align}
and the inner product between two waveform models is defined as

\begin{align}
\label{}
(\tilde{h}_1 \mid \tilde{h}_2) \equiv 2 \int_{f_{low}}^{f_{upper}} \frac{\tilde{h}_1 \tilde{h}_2^{*} + \tilde{h}_1^{*} \tilde{h}_2}{S_n(f)} df.
\end{align}

In the inner product, the superscript star stands for complex conjugation, and $S_n(f)$ is the detector spectral noise density. With this definition of the inner product, the SNR is defined as
\begin{align}
\label{SNR}
\rho^2 \equiv (\tilde{h} \mid \tilde{h}) = 4 Re \int_{f_{low}}^{f_{upper}} \frac{\mid \tilde{h}(f)\mid^2}{S_n} df.
\end{align}

\begin{figure}
\includegraphics[width=3.5in,height=2.5in]{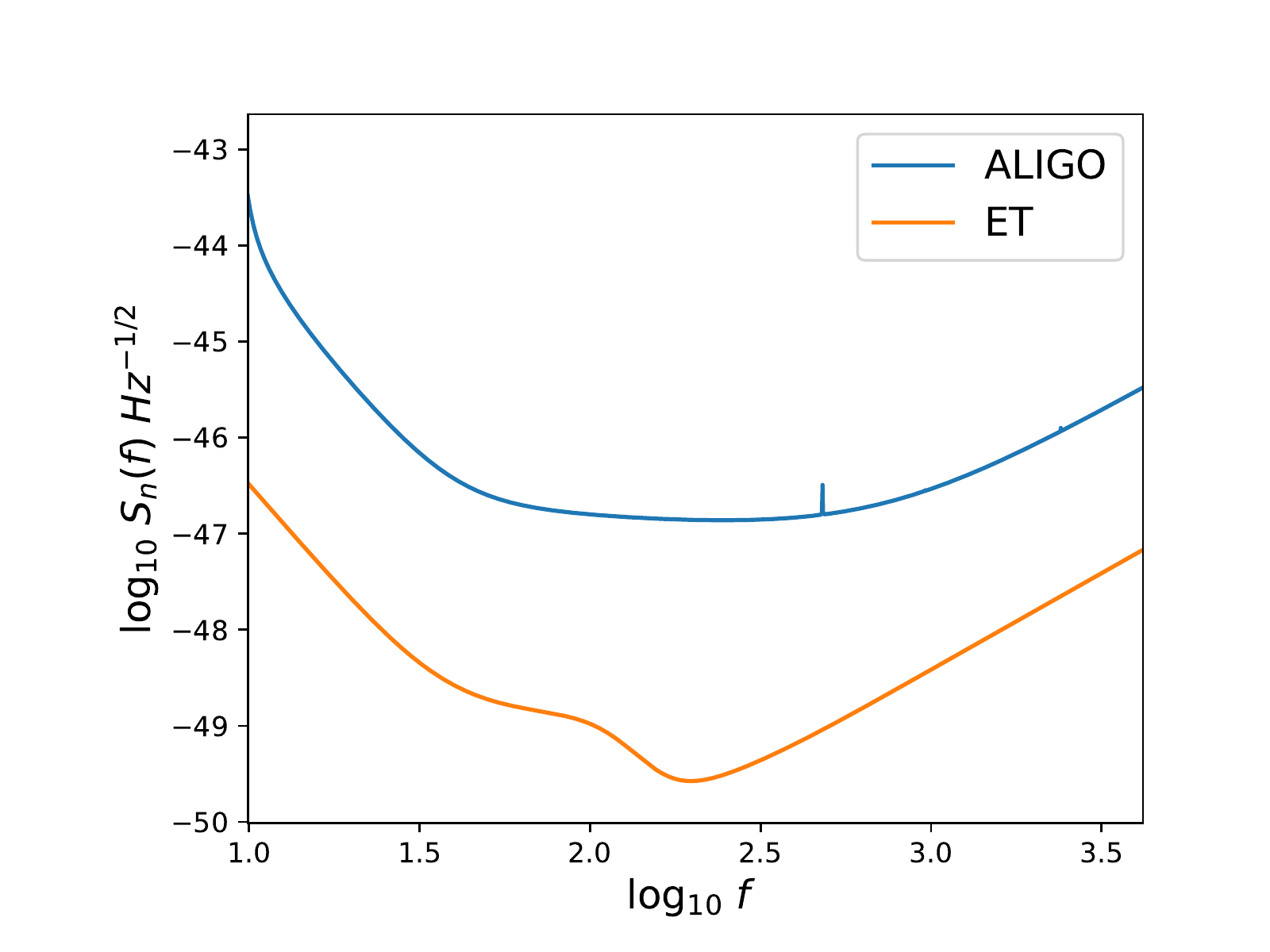}
\caption{Detector spectral noise density for the Advanced LIGO (aLIGO) and Einstein Telescope (ET).}
\label{Sn}
\end{figure}

In order to move on we need to consider a GW form. For our purposes, let us take a post-Newtonian waveform known as TaylorF2, that uses the stationary phase approximation for the waveform at 3.5PN expression for the orbital phase of inspiraling binary system with aligned spins \cite{waveform01, waveform02, waveform03, waveform04, waveform05, waveform06}. For a coalescing binary with component masses $m_1$ and $m_2$, in the frequency domain and in the angle-averaged approximation, the gravitational waveform reads
\begin{align}
\label{waveform}
\tilde{h}(f) = \mathcal{A} f^{-7/6} e^{i\Phi(f)},
\end{align}
where $\mathcal{A}$ is the amplitude, which takes the form

\begin{align}
\label{A}
\mathcal{A} \propto \mathcal{M}^{5/6}_c/d_L^{GW},
\end{align}
and to 3.5 PN order the phase waveform $\Phi(f)$ is given by
\begin{align}
\label{phi}
\Phi(f) = 2 \pi f t_c - \phi_c - \frac{\pi}{4} + \frac{3}{128 \eta v^5} \Big[ 1 + \sum_{i=2}^7 \alpha_i v^i  \Big],
\end{align}
where the coefficients $\alpha_i$ are the corrections up to 3.5 PN order, which includes the spin parameters $\chi_1$ and $\chi_2$ of each component of the binary system (for the definition of the coefficients see, e.g., \cite{nitz13}). Other parameters are $v = (\pi M f)^{1/3}$, $M = m_1 + m_2$, $\eta = m_1 m_2/(m_1 + m_2)^2$, and $\mathcal{M}_c = (1+z) M \eta^{3/5}$ as being the inspiral reduced frequency, total mass, symmetric mass ratio, and the redshifted chirp mass, respectively. The quantities $t_c$ and $\phi_c$ are the time and phase of coalescence, respectively. In order to simplify the analysis, the positions of the sources in the sky are assumed to be known and, hence, in what follows the angles in the celestial sphere do not enter as parameters. Therefore, our baseline parameter model is given by
\begin{equation}
\label{baseline1}
\begin{aligned}
\theta_i = \{M, \eta, \chi_1, \chi_2, t_c, f_c, b \},
\end{aligned}
\end{equation}
where $b$ is the parameter of the model that characterizes the deviations with respect to the GR theory. 

Assuming Eq. (\ref{dL_Gw_ft}) and Eq. (\ref{A}), one can promptly see that for $\beta_T = 0$ the GW amplitude from inspiraling of compact binary systems in the GR is recovered, as expected. In general, the modified GW form for binary systems in the inspiral phase is completely determined by the combination of Eqs. (\ref{dL_Gw_ft}) - (\ref{baseline1}). In what follows, we will be concerned only with modifications due to the $f(T)$ gravity in the inspiral part of the waveform. Also, a similar methodology has been adopted using the parameterized post-Einsteinian framework \cite{ppE01, ppE02}.

\section{Forecasts for ground-based interferometers}
\label{results}

In this section, we investigate the consequences of the above modified GW form propagation obtained in the context of the $f(T)$ gravity, focusing on forecasts for two specific ground-based GW detectors, namely, aLIGO \cite{ALIGO} and ET \cite{ET02}. Figure \ref{Sn} shows $S_n$ for both detectors. 

In what follows, we apply the Fisher information, as presented in the last section, to estimate new constraints on $b$ by means of  gravitational waves from compact binary coalescences. It is worth stressing that such a procedure has never been done before in the literature.

In our simulations, we consider three types of compact binary systems, applied to both detectors. 
\\

1 - Black hole $-$ black hole (BBH). The black hole masses are chosen to be uniform in the interval $[10 - 30]~ {\rm M}_{\odot}$ under the condition $m_1 \gtrsim m_2$ and $\eta < 0.25$. The spin magnitudes $\chi_1, \chi_2$ associated with each mass, are chosen to be uniform in the interval $[-1, 1]$.
\\
\\
2 - Neutron star $-$ neutron star (BNS). The distribution of the neutron star masses is chosen to be uniform within $[1 - 2]~{\rm M}_{\odot}$, also under the condition $m_1 \gtrsim m_2$ and $\eta < 0.25$. In this case, in the mock data generation let us take $\chi_1 = \chi_2 = 0$.
\\
\\
3 - Black hole $-$ neutron star (BBHNS). The black hole mass is chosen to be uniform in the interval $[10 - 30]~ {\rm M}_{\odot}$ with $\chi_{BH}$ chosen to be uniform in the interval $[-1, 1]$. For the NS, let us take the range of masses $[1 - 2]~{\rm M}_{\odot}$ with $\chi_{NS} = 0$.
\\

The BHs mass range is chosen to be compatible with the masses of the BHs already detected and cataloged by LIGO/VIRGO team \cite{LIGO01}. Also, these values are compatible with the frequency range from which will be possible to observe such binary systems with ET \cite{ET_design01,ET_design02}. The NS mass range is fully compatible with theoretical models and NS observations.

The redshift distribution of the sources is taken to be of the form \cite{Pz_merger}
\begin{equation}
P(z) \propto \frac{4 \pi d^2_C(z) R(z)}{H(z)(1+z)},
\end{equation}
where $d_C$ is the comoving distance and $R(z)$ describes the time evolution of the burst rate and is defined to be: $R(z) = 1 + 2z$ for $z \leq 1$, $R(z) = 3/4(5 - z)$ for $1 < z \leq 5$ and null for $z > 5$.

The generation of catalogs is carried out as follows. We first simulate the redshift measurements according to the redshift distribution. At every simulated redshift, we randomly sample the mass and spins of the BH and NS according to the above specifications. Then, we calculate the SNR of each event and confirm that it is a GW detection if $\rho > 8$ (SNR $>$ 8). For every confirmed detection, we calculate the modified wave form and the inverse of the Fisher matrix in order to estimate the borders of our free baseline parameter. In all cases, unless stated otherwise, we consider $b = 0$ as the injection value to run the analysis. When performing the integration for aLIGO we assume the entire frequency band on $S_n$ \cite{ALIGO}. For ET, we assume $f_{low} = 1$ Hz and $f_{upper} = 2 f_{LSO}$, where $f_{LSO} = 1/(6^{3/2} 2 \pi M_z)$ is the orbital frequency at the last stable orbit \cite{Pz_merger}, with $M_z = (1 + z)M$. Also, let us take $\Omega_{m0} = 0.308$ and $H_{0} = 67.8$ km/s/Mpc \cite{Planck}. In what follows, we present our main numerical results.

\begin{figure*}
\includegraphics[width=2.2in,height=2.0in]{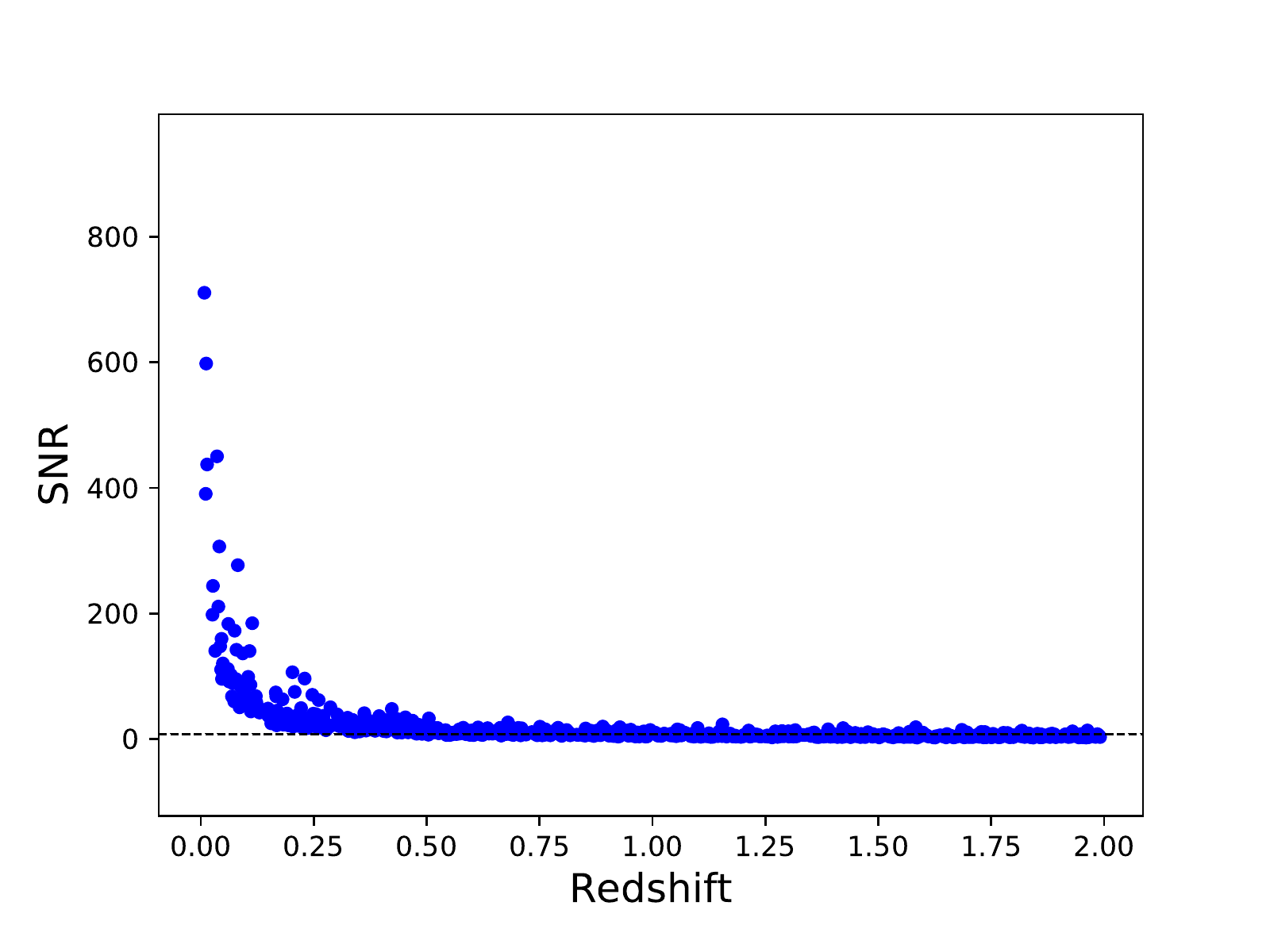} \quad  
\includegraphics[width=2.2in,height=2.0in]{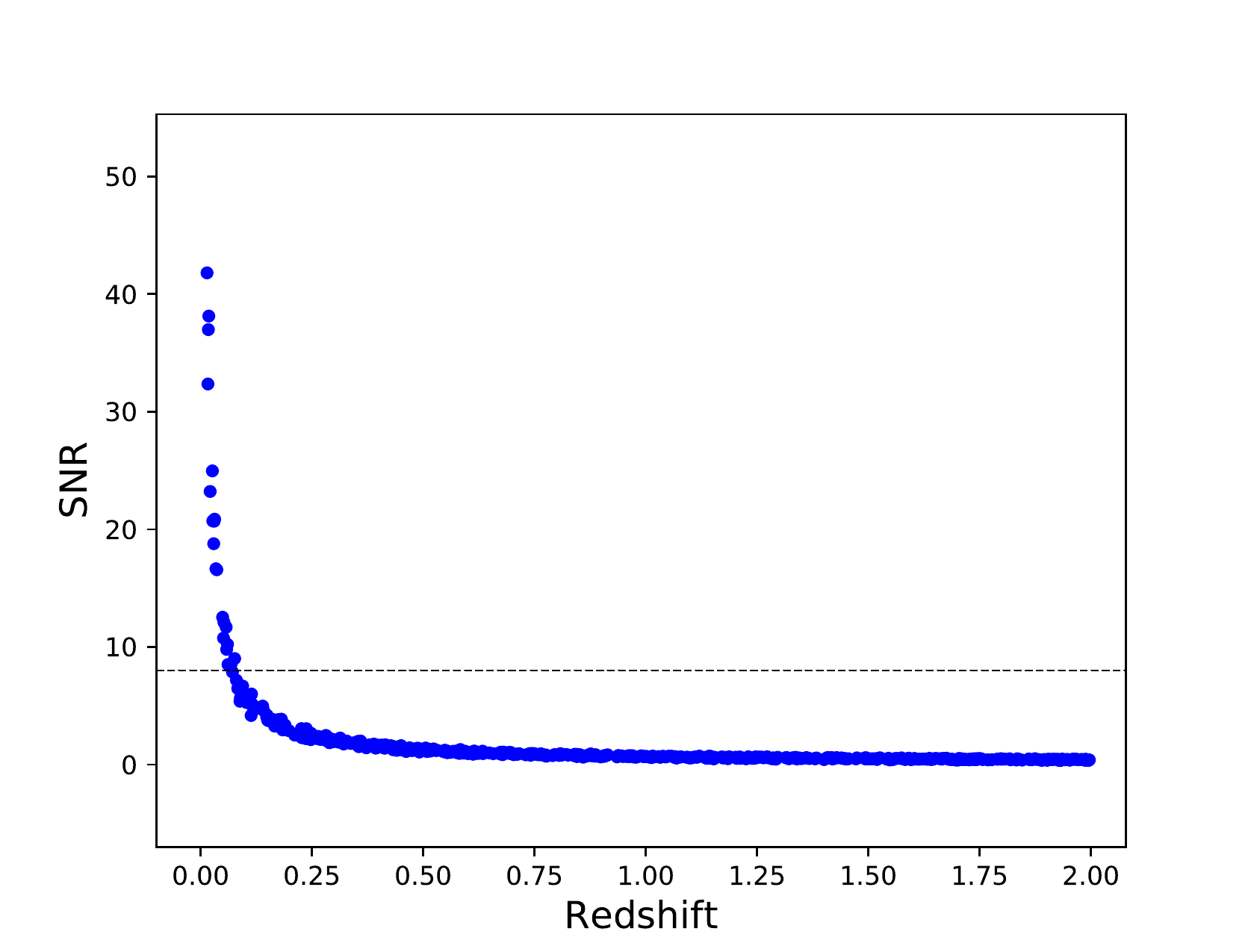} \quad
\includegraphics[width=2.2in,height=2.0in]{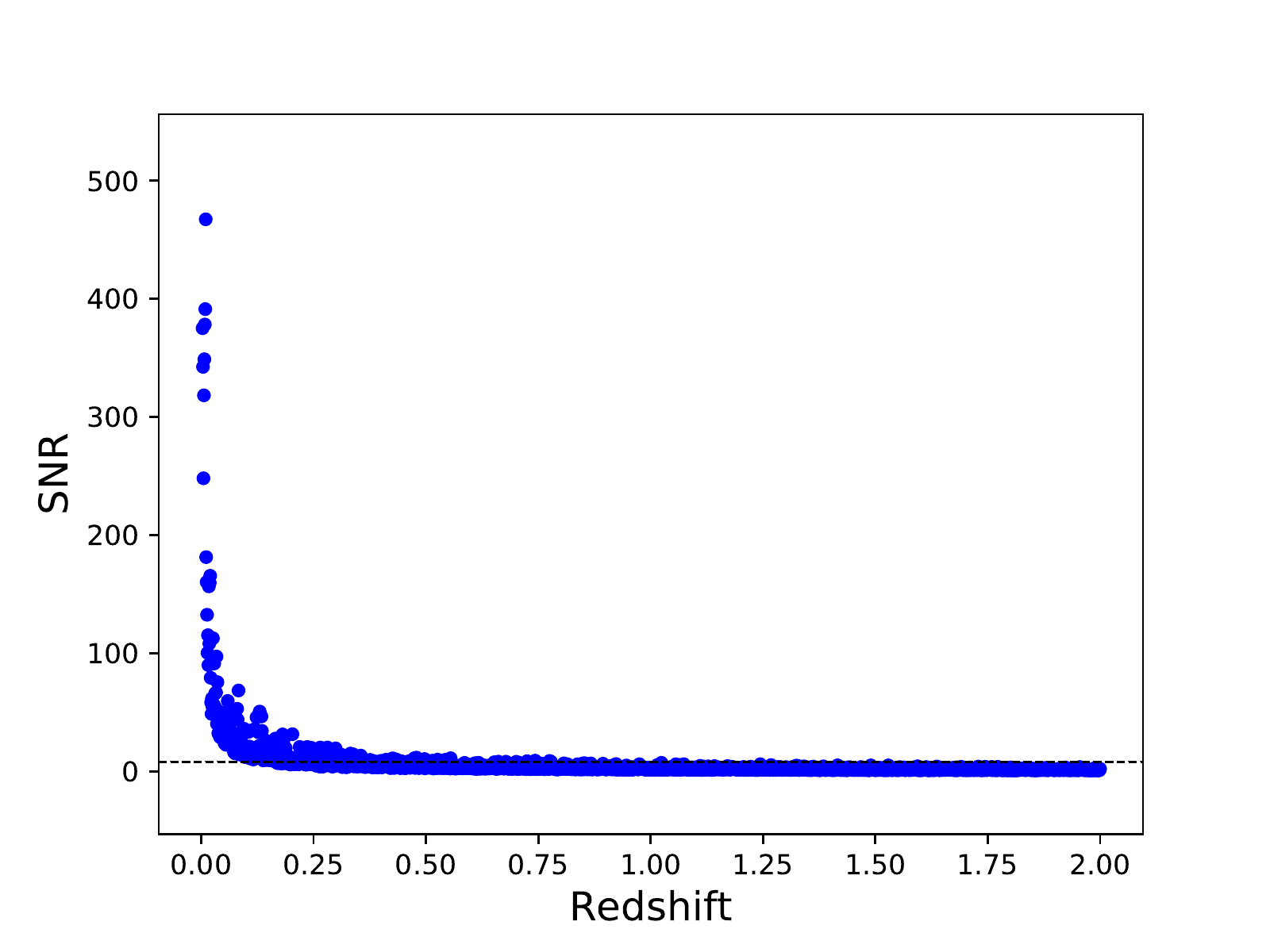} 
\caption{Left panel: Signal-to-noise ratio (SNR) as a function of $z$ for several BBH simulated events assuming Advanced LIGO noise power spectral density sensibility.
Middle panel : Same as left panel, but for BNS events. Right panel: Same as left panel, but for BBHNS events. In all panels the black lines represent SNR = 8.}
\label{SNR_aligo_BH}
\end{figure*}

\begin{figure*}
\includegraphics[width=3.0in,height=2.5in]{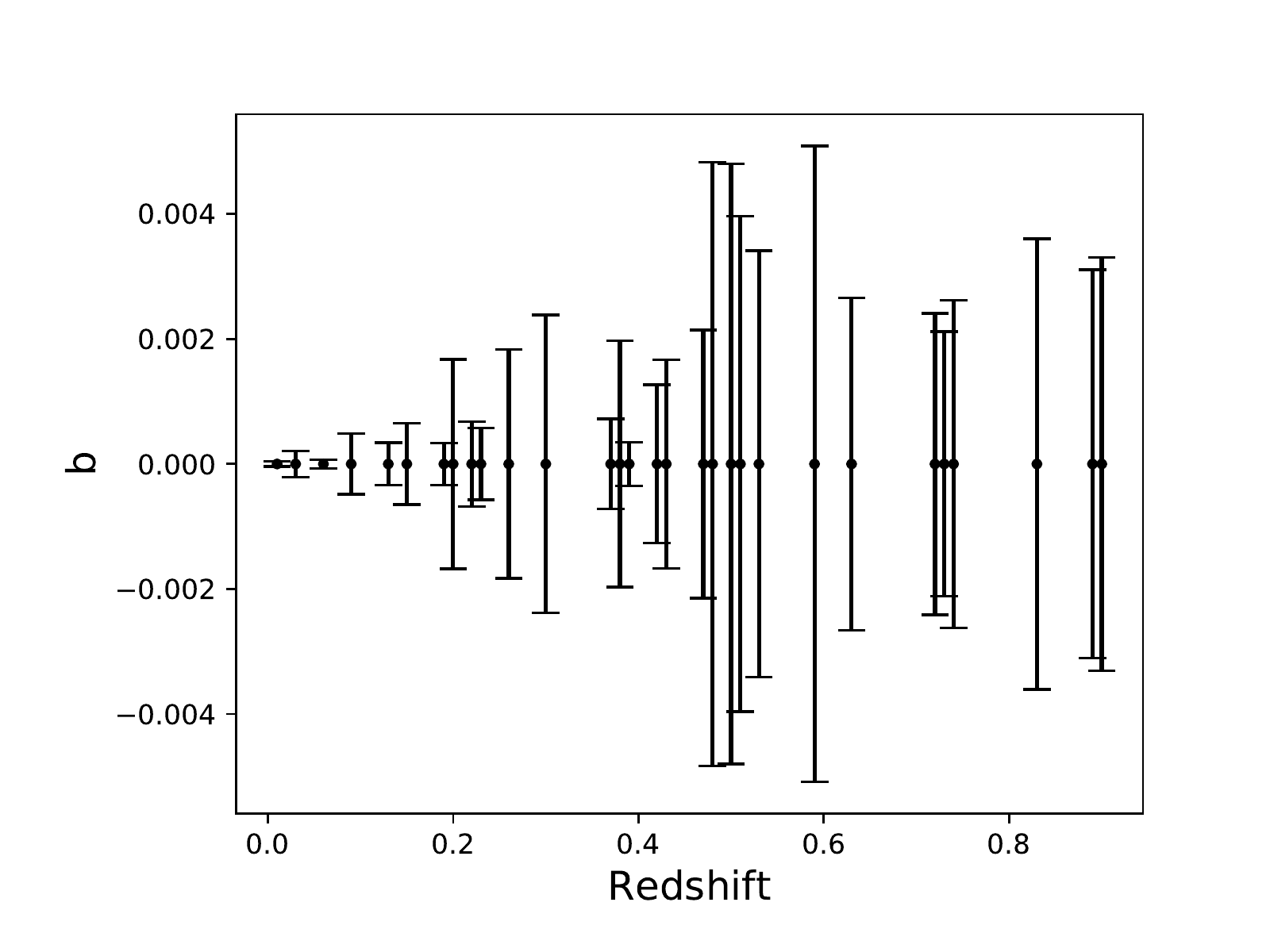} \quad 
\includegraphics[width=3.0in,height=2.5in]{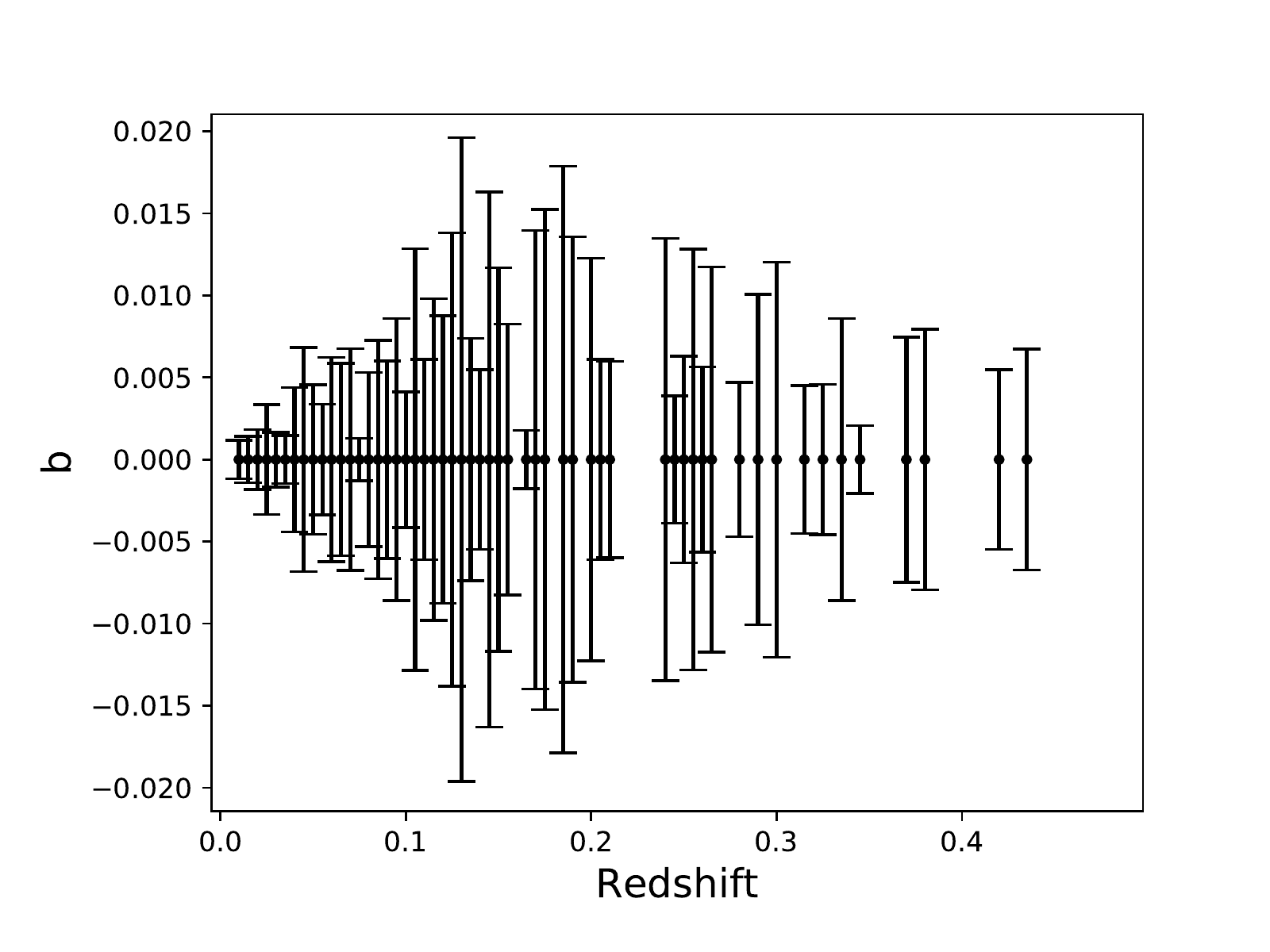} \quad 
\caption{Left panel: Estimates on $b$ at 68\% CL assuming BBH events on every specific redshift within range $z \in [0, 1]$ from the perspective of the aLIGO power spectral density sensitivity. Right panel: Same as left panel, but for BBHNS events up to the $z$-limit where such events can be detected.}
%Joint analysis in the parameter space $b - M$ from all events shown in the left panel.}
\label{aligo_b_results}
\end{figure*}

%\begin{figure*}
%\includegraphics[width=3.0in,height=2.5in]{aligo_SNR_z_BHNS.pdf} \quad \quad 
%\includegraphics[width=3.0in,height=2.5in]{aligo_BHNS_b.pdf}
%\caption{Left panel: Signal-to-noise ratio (SNR) as a function of the $z$ for several BBHNS simulated events assuming Advanced LIGO noise power spectral density sensibility. Right panel: Estimates on $b$ assuming events on every specific redshift within range $z \in [0, ]$.}
%\label{aligo_BHNS01_results}
%\end{figure*}

It is worth emphasizing that we are deriving constraints (forecast analysis) on $b$ directly evaluating how the gravitational signal $\tilde{h}(f)$ is modified due to the expansion of the Universe in the context of the $f(T)$ gravity, when the signal is generated by the source until it reaches the GWs detector. On the other hand, the most common procedure in the literature is to assume GW sources to determine the luminosity distance, as standard sirens, associated with each event, and then using the luminosity distance mock events to get constraints on the cosmological parameters. It is important to realize that we are not following this procedure. However, notice that within the approach used here, i.e., the evaluation of the free parameter of the theory directly by the GW signal/strain, one has stronger constraints in the case of few detected events, such as the case of simulations using the aLIGO sensitivity. Hence, it is not only more reasonable, but it seems also more promising to investigate deviations from GR by assuming directly modifications on $\tilde{h}(f)$ in the statistical forecast (see \cite{ppE01, ppE02} and references therein for a similar methodology as that applied here, but within the parameterized post-Einsteinian framework context).

In what follows, we present and discuss our statistical results.

\subsection{Advanced LIGO}

In this subsection, we present our results considering the aLIGO detector. As an example, in the left panel of Figure \ref{SNR_aligo_BH} we show a mock catalog with several BBH events assuming the noise power spectral density for aLIGO. The redshift of each source is generated between 0 and 2. In the left panel, we show the SNR associated with each event as a function of the redshift. The black line corresponds to SNR = 8. As expected, we can see SNR $\propto z^{-1}$, and we note than for $z \gtrsim 1.5$ most of the sources have SNR $<$ 8. 

Similarly, in the middle panel, we show a mock catalog for BNS events. For $z \gtrsim 0.13$, the sensitivity of the aLIGO is not high enough to detect BNS events with SNR $>$ 8. Thus, GWs from BNS events from our mock data can only be observed at very low redshifts. Recall that the GW170817 event ocurred at $z \simeq 0.009$ with SNR $\simeq$ 33. Our mock sources presented very similar values when evaluated at $z << 1$. 

Finally, in the right panel, we show the SNR as a function of $z$ for mock catalog of BBHNS events. In this case, most of the GW sources with $z \gtrsim 0.4$ have SNR $<$ 8.

In order to impose new constraints on $b$, we set the network SNR threshold to be a real detection to SNR = 8 and only sources with SNR $>$ 8 are kept in our analysis. Therefore, when using the aLIGO sensitivity, the BNS events are not taken into account to constrain the $b$ parameter, once only sources at $z << 1$ present SNR $>$ 8 as noticed above. This is because it is necessary to have intermediary and high redshift observations in order to probe modifications in the gravitational theory.

The left and right panel of Figure \ref{aligo_b_results} shows estimates/borders on $b$ at 68\% confidence level (CL) for a BBH (BBHNS) mock catalog for $0 < z < 1$ ($0 < z < 0.5$), on each specific $z$, assuming only events with SNR $>$ 8. It is important to keep in mind that we are conducting a forecast analysis. Thus, only upper (lower) values on the free parameter of the theory, centered on $ b = 0 $ (GR), are the outputs from the Fisher information. It is very important because we will have an estimate of what the maximum sensitivity constraints on the theory (in our case $ b $) will be in the future.

For comparison, let us take some recent estimates on the parameter $b$ using cosmological probes. In \cite{fT09,fT10} a joint analysis with geometric data show that $b \sim \mathcal{O}(10^{-2})$, in \cite{fT11} a robust analysis using CMB anisotropy data show that $b \sim \mathcal{O}(10^{-3})$. Therefore, notice that the constraint via GWs on $f(T)$ gravity within aLIGO sensitivity will be similar to those imposed via cosmological probes.

\begin{figure*}
\includegraphics[width=2.2in,height=2.0in]{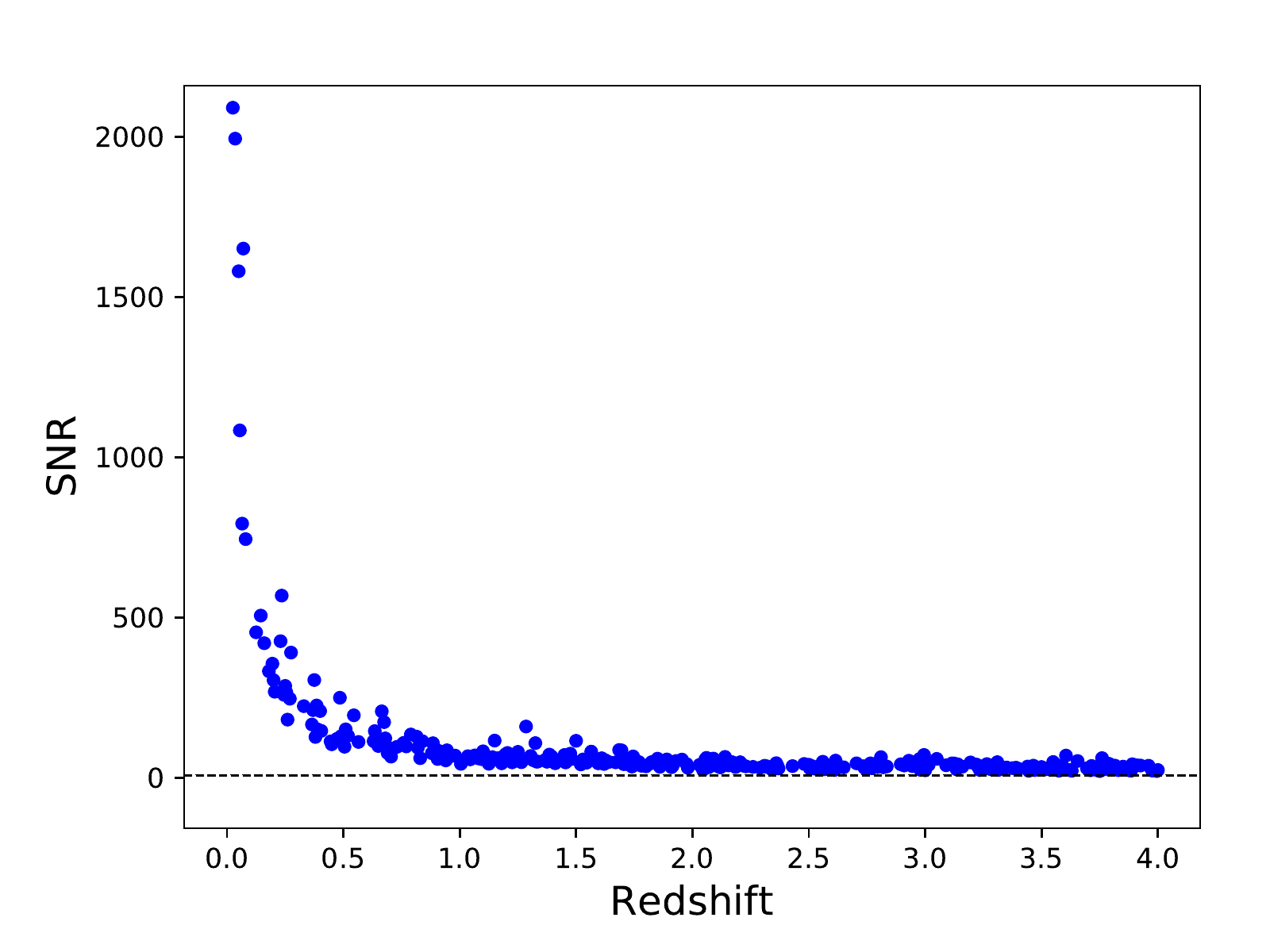} \quad 
\includegraphics[width=2.2in,height=2.0in]{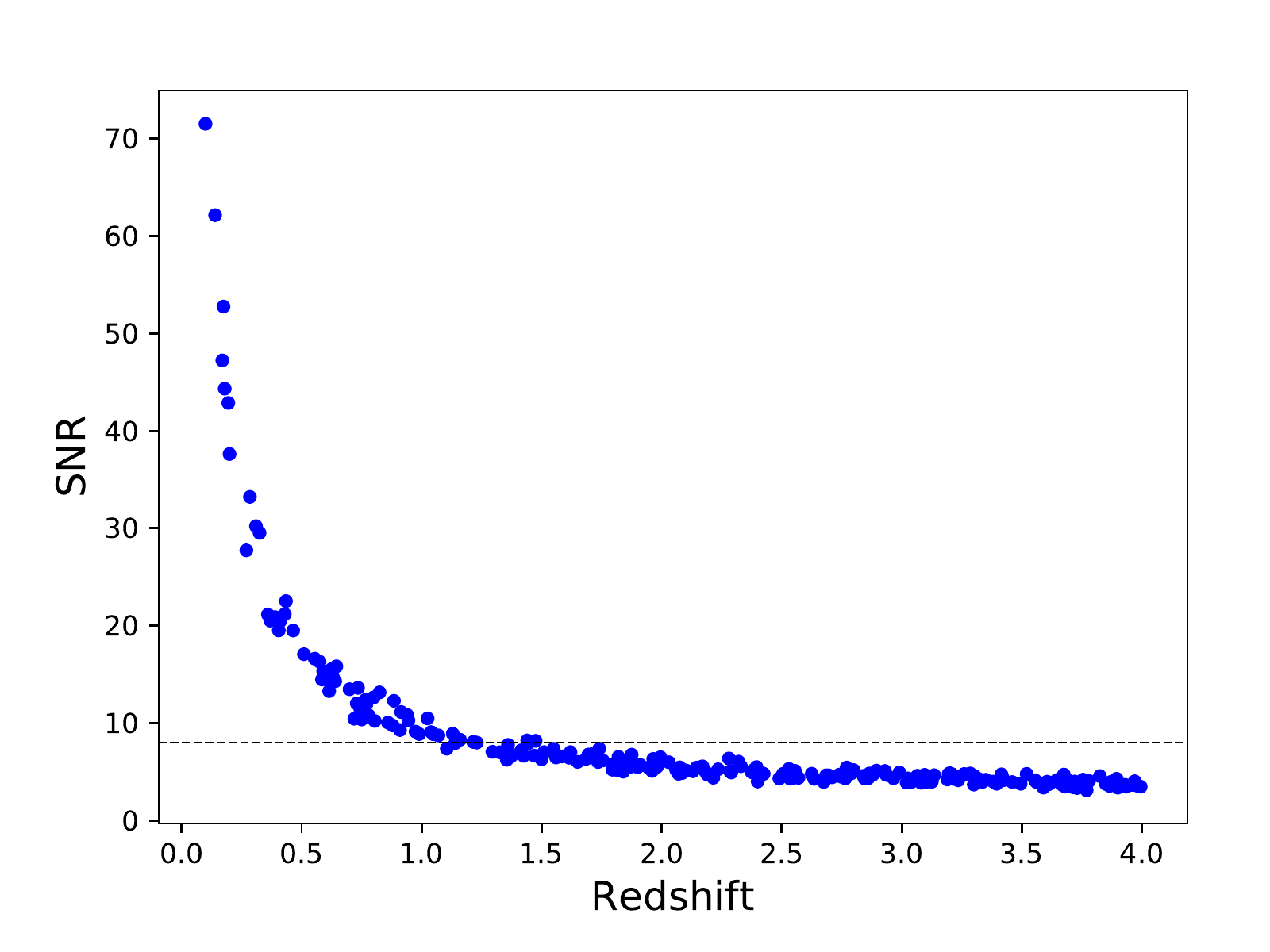} \quad
\includegraphics[width=2.2in,height=2.0in]{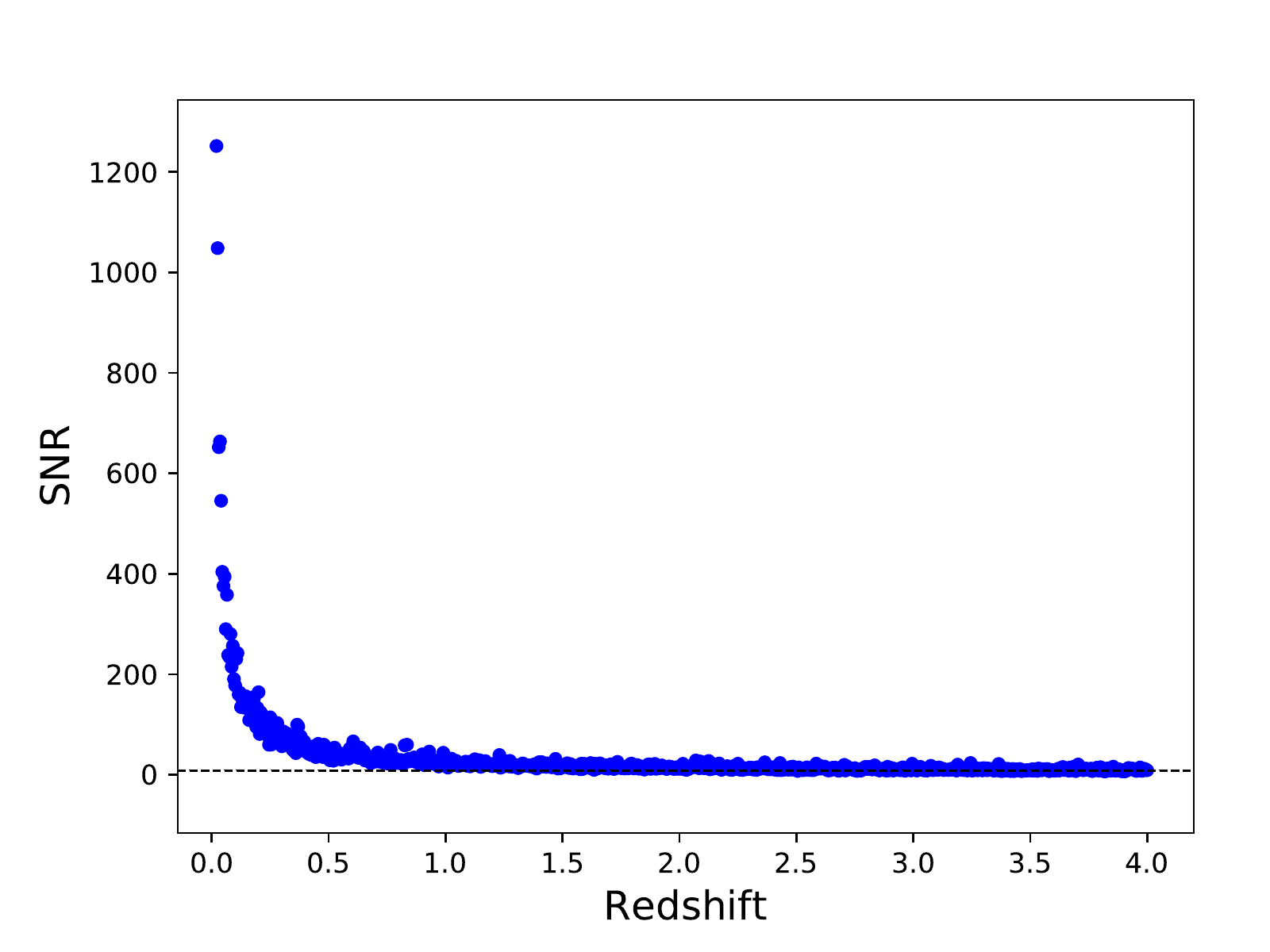} 
\caption{Left panel: Signal-to-noise ratio (SNR) as a function of the $z$ for several simulated BBH events assuming the ET spectral density sensibility. Middle panel : Same as left panel, but for BNS events. Right panel: Same as left panel, but for BBHNS events. In all panels the black line represents SNR = 8.}
\label{SNR_ET}
\end{figure*}

\begin{figure*}
\includegraphics[width=2.2in,height=2.0in]{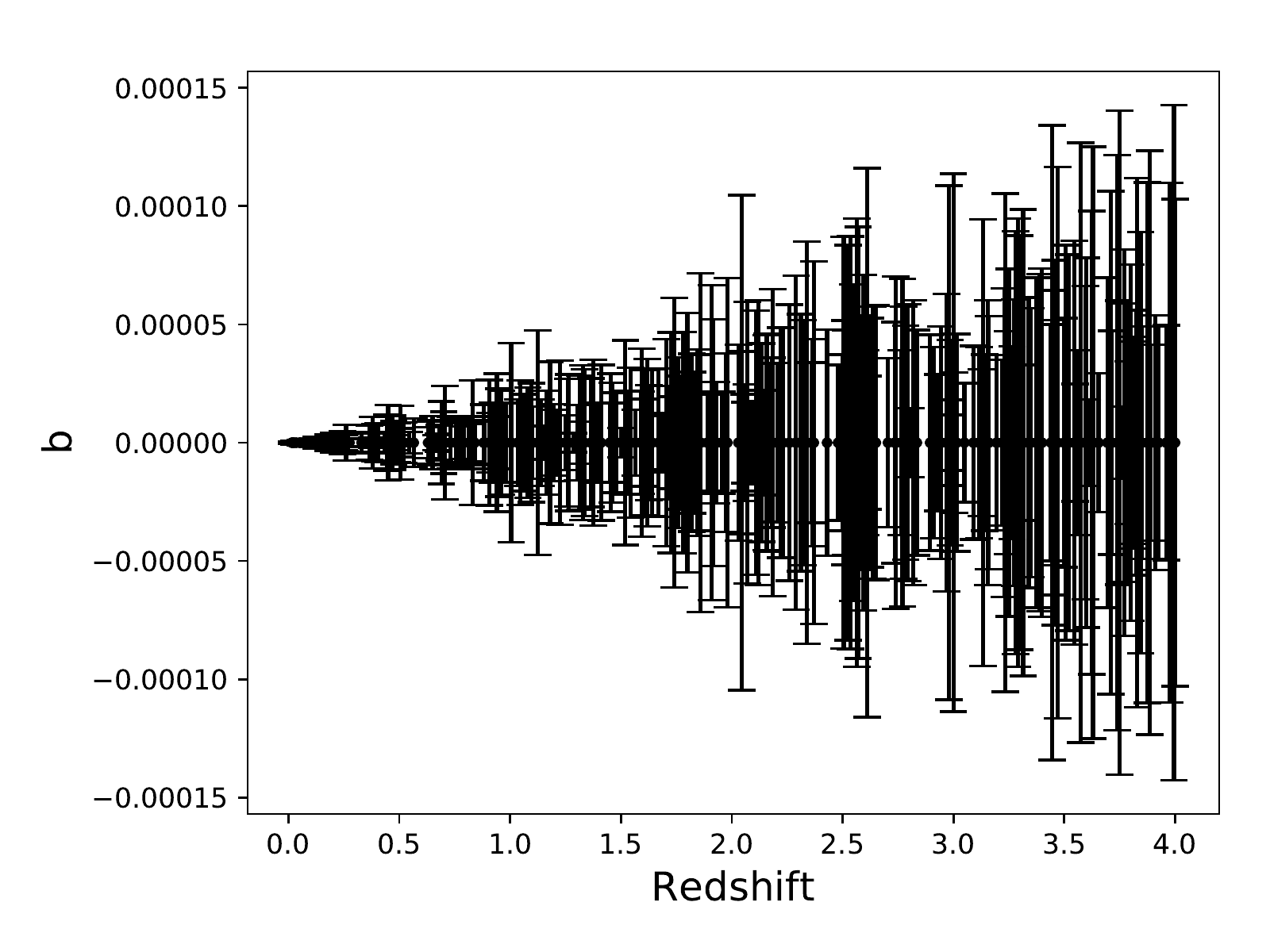} \quad
\includegraphics[width=2.2in,height=2.0in]{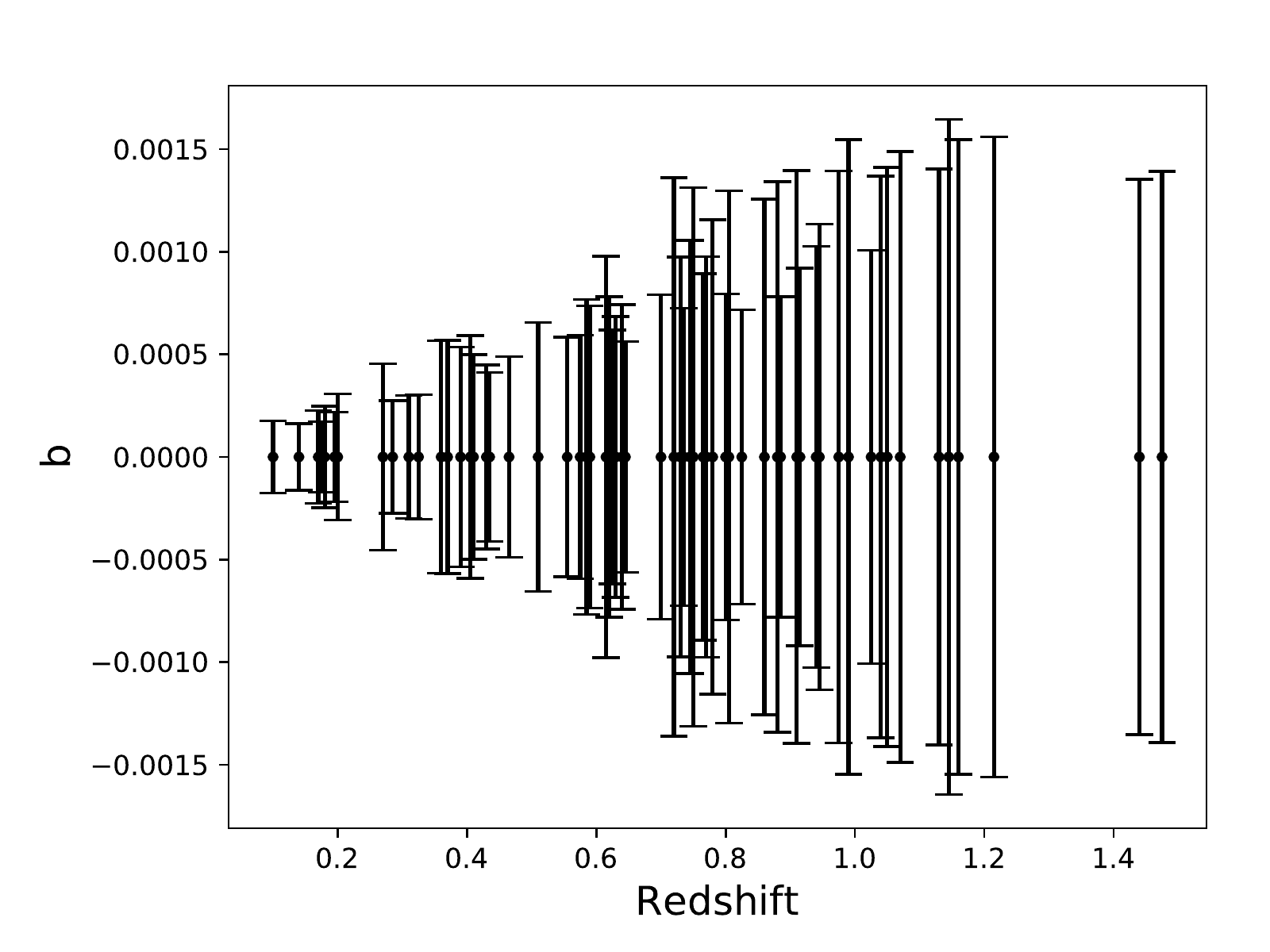} \quad 
\includegraphics[width=2.2in,height=2.0in]{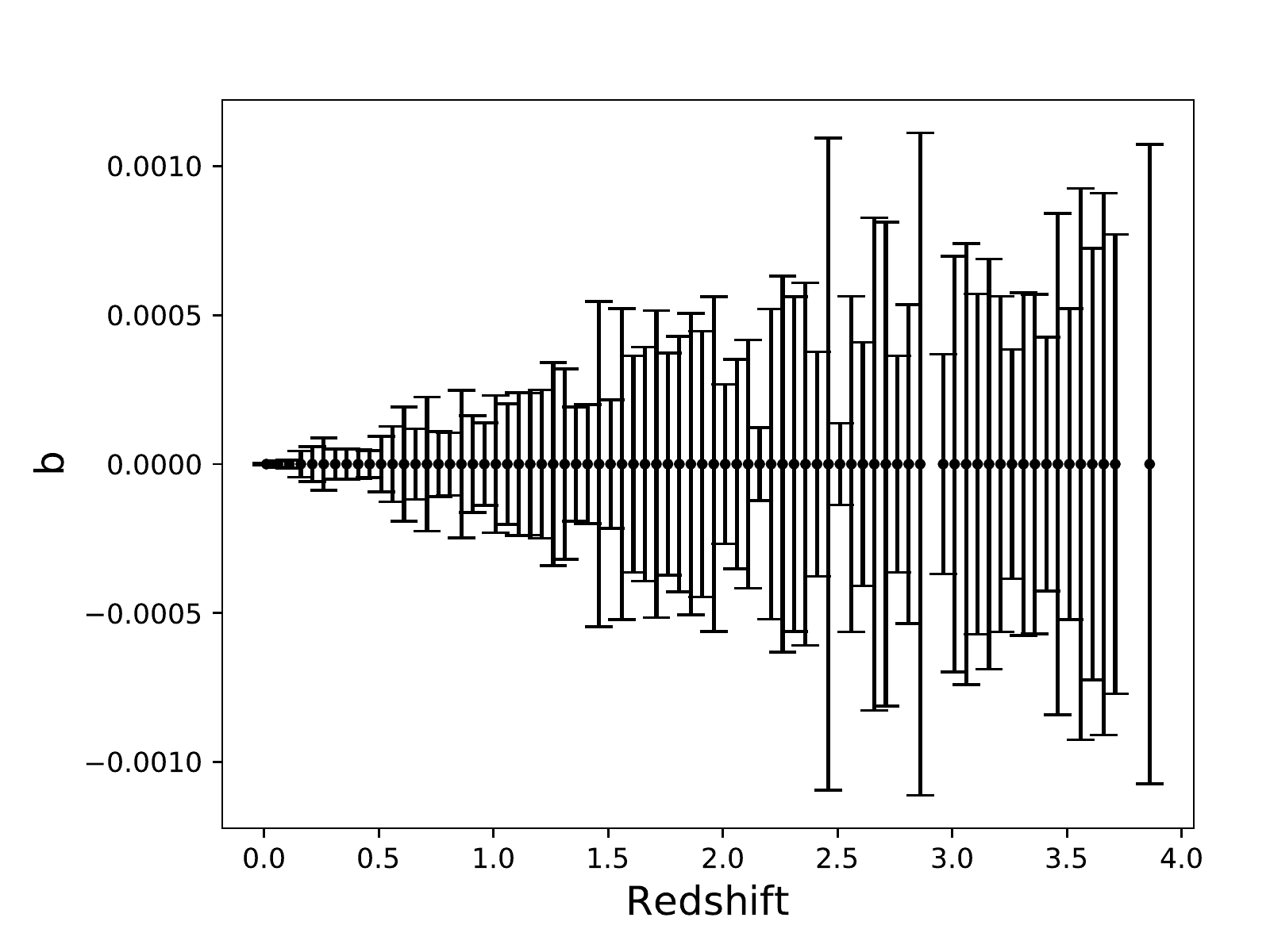} 
\caption{Left panel: Estimates on $b$ at 68\% CL assuming several BBN events on every specific redshift within range $z \in [0, 4]$ from the perspective of the ET power spectral density sensitivity. Middle panel : Same as left panel, but for BNS events. Right panel: Same as left panel, but for BBHNS events.}
\label{ET_b}
\end{figure*}

\subsection{Einstein Telescope}

In this subsection, we present our results now considering the ET detector. The ET is a third-generation ground-based detector of GWs and it is envisaged to be ten times more sensitive in amplitude than the advanced ground-based detectors in operation, covering the frequency band ranging from 1 Hz to $10^4$ Hz. 

Unlike current detectors, from ET conceptual design study, the expected rates of BNS detections per year are of the order of $10^3-10^7$~\cite{ET02}. However, we can expect only a small fraction ($\sim 10^{-3}$) of them accompanied with the observation of a short $\gamma$-ray burst. If we assume that the detection rate is in the middle range around $\mathcal{O}(10^5)$, we can expect to see $\mathcal{O}(10^2)$ events with the short $\gamma$-ray burst per year. Therefore, we can simulate 100 - 1000 BNS events, as being a reasonable number of events.

Since the ET is composed of three independent interferometers (not correlated), the combined SNR can be written as
\begin{equation}
\rho = \sum_{i=1}^3 (\rho^{(i)})^2.
\end{equation}

As an example, the left panel of Figure \ref{SNR_ET} shows a mock catalog of  BBH events assuming the ET noise power spectral density. In the middle panel, we show a mock catalog considering BNS sources. Here, it can be clearly seen how much the ET is more sensitive than aLIGO, especially for the BNS sources. Using ET, we can have BNS events, i.e., sources with SNR $>$ 8, up to $z \simeq 1.5$. BBH events can be observed up to high $z$ values with a significant high value of SNR. Thus, with ET we have the opportunity to detect (or simulate) sources with larger $z$, including BNS sources. On the right panel, we show SNR vs. $z$ for BBHNS mock events, and we find that SNR $>$ 8 is also predominate up to high $z$ like BBH events, but for BBHNS the SNR values decays faster and present lower SNR values when compared to BBH events. Notice also that now the BNS sources (and BBHNS sources) can cross the barrier $z \simeq 0.60$. Thus, it is possible to go to redshifts for which the Universe experiences the transition between the decelerated and accelerated phases. It is also evident that by assuming ET we have a bigger SNR associated with the events, being possible to estimate the parameters with greater precision.

The left panel of Figure \ref{ET_b} shows estimate sensitivity on $b$ at 68\% CL for a mock BBH source between $z = 0$ up to $z = 4$. Similarly, we show also in the middle and right panels the BNS and BBHNS sources. Notice that now there is enough sensitivity to observe at high redshifts. As noticed before, and shown in Figure \ref{SNR_ET}, a GW detection of BNS sources is very difficult for $z \gtrsim  1.5$. For all the three cases, the increasing of the error bars with redshift is explained by the decrease of the SNR, which is inversely proportional to the redshift. Therefore, for the ET detector, the estimates on $b$ are $\mathcal{O}(10^{-4})$, $\mathcal{O}(10^{-3})$ and $\mathcal{O}(10^{-3})$ for BBH, BNS and BBHNS, respectively. On the other hand, if we consider only $z << 1$, we obtain $b \sim  \mathcal{O}(10^{-5})$ for BBH sources and $\mathcal{O}(10^{-4})$  for BBHNS. In comparison with the recent estimatives on $b$ \cite{fT09, fT10, fT11}, we have that BBH and BBHNS sources can improve current estimates up to 2 orders of magnitude and BNS sources presented very similar results when compared to BBH source from aLIGO. For a concrete comparison, if one observes the bounds $b \sim \mathcal{O}(10^{-5}) - \mathcal{O}(10^{-4})$ from future GW events, and we take the current best fit values on $b$, the estimates sensitivity may be accurate enough to probe $f(T)$ gravity at $\sim$ 99\% CL beyond GR. 

\subsection{Comparison with GR}

\begin{figure}
\includegraphics[width=3.5in,height=2.5in]{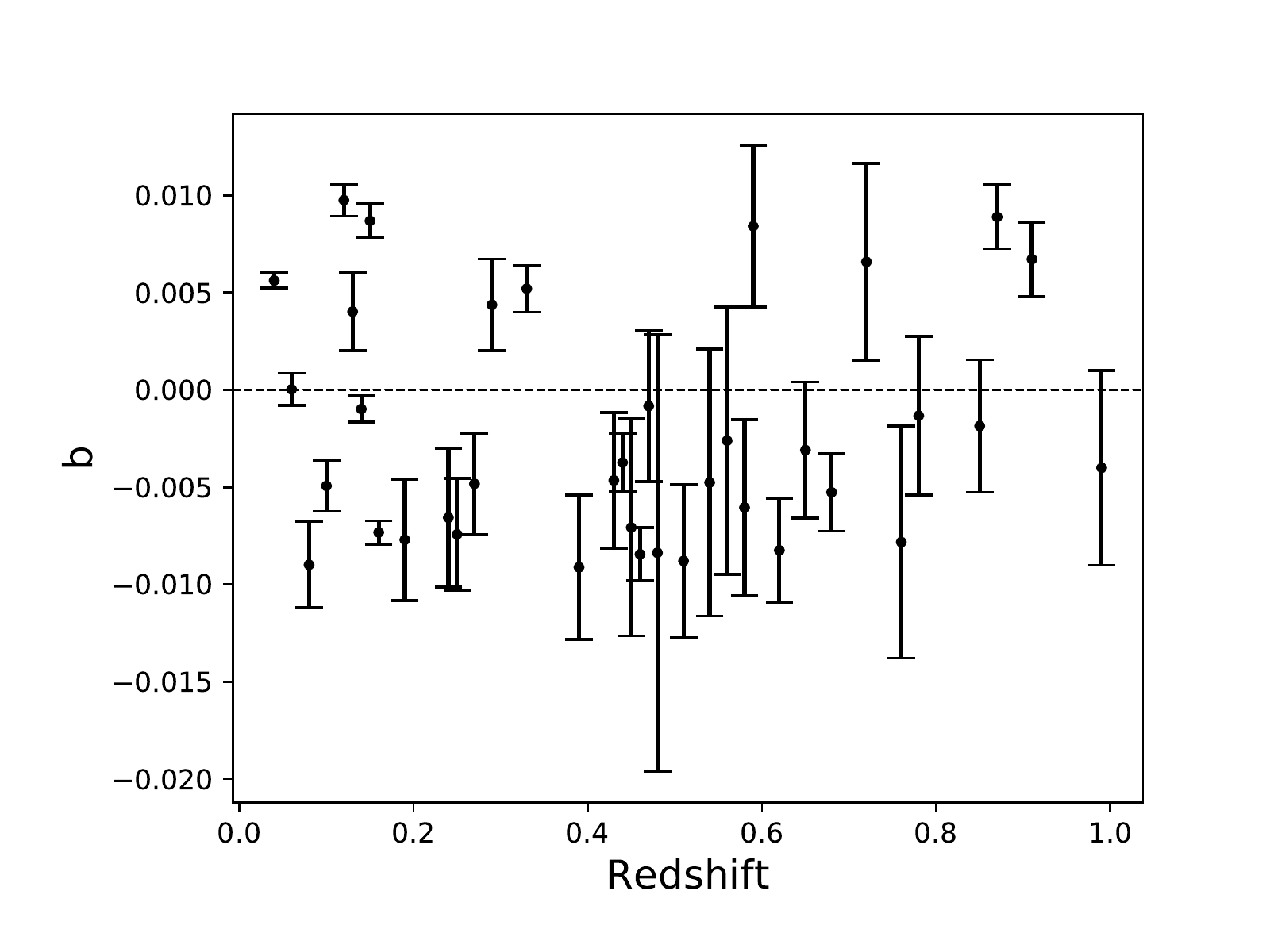} \quad 
\caption{Estimates at 68\% CL on $b$ from a randomly $b$ sample assuming BBH source on every specific redshift within aLIGO power spectral density noise. All events have SNR $>$ 8. The black solid line at $b = 0$ corresponds to GR.}
\label{aligo_prior_b}
\end{figure}

\begin{figure*}
\includegraphics[width=2.2in,height=2.0in]{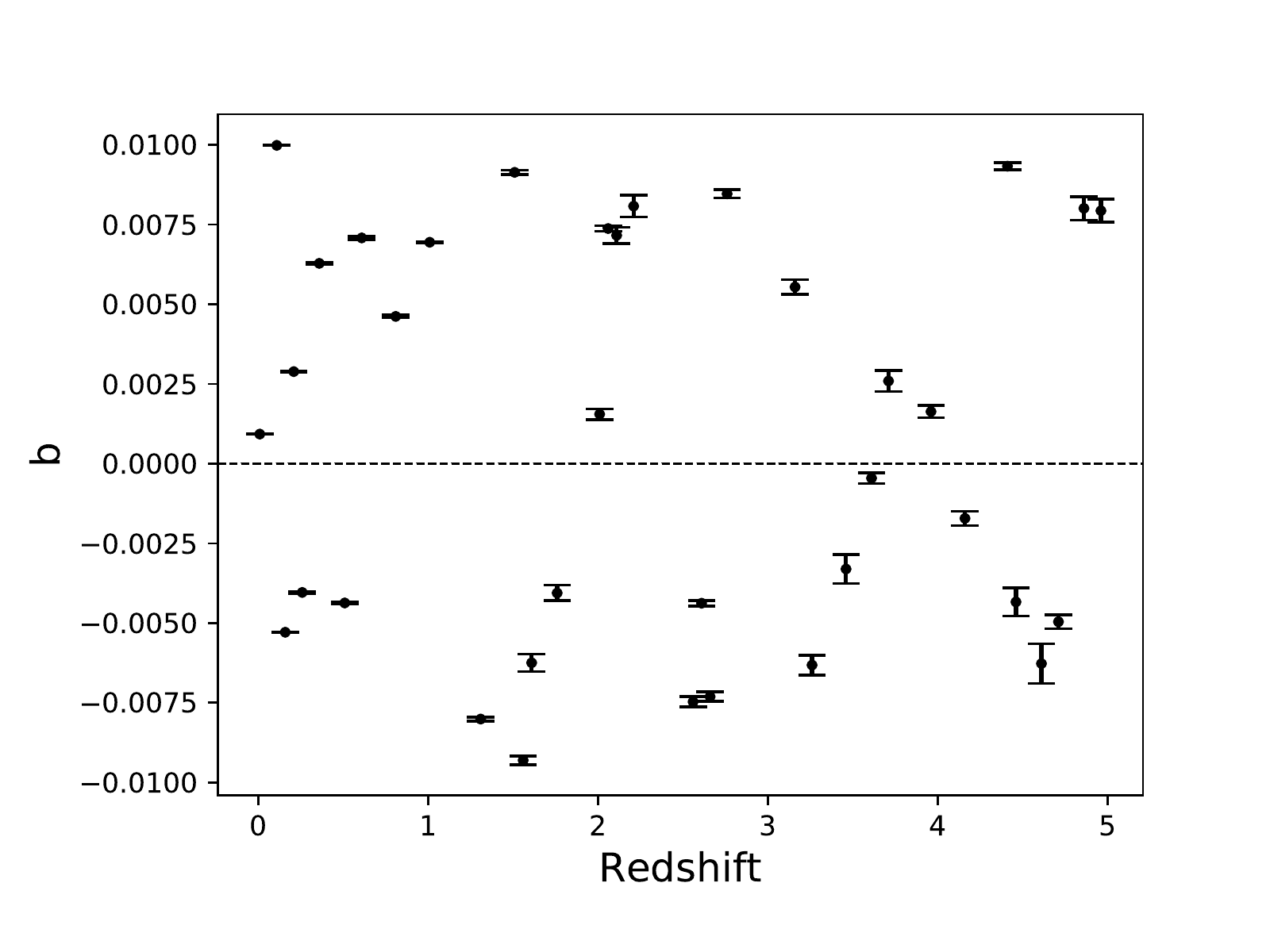} \quad
\includegraphics[width=2.2in,height=2.0in]{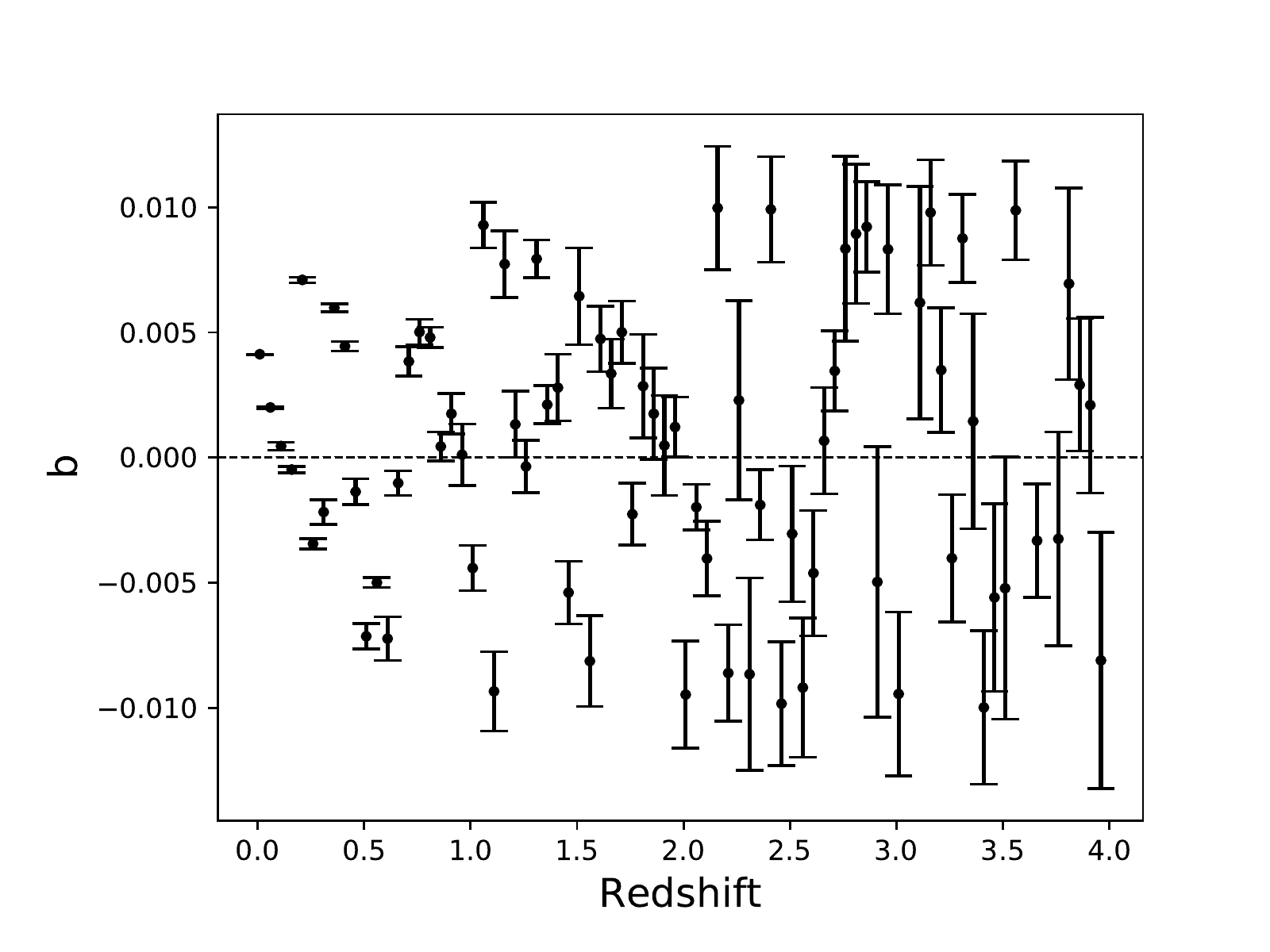} \quad
\includegraphics[width=2.0in,height=2.0in]{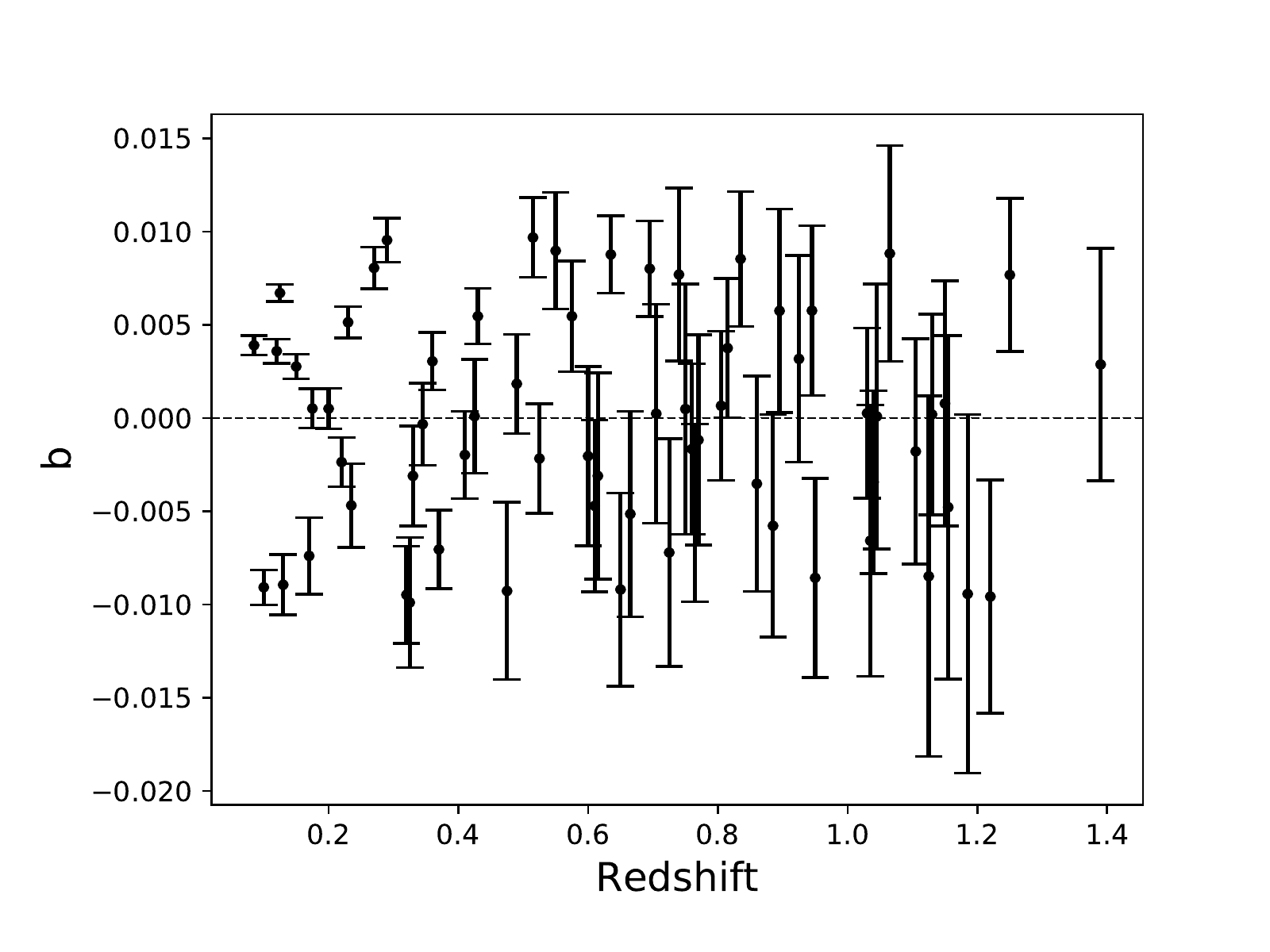}
\caption{Left panel: Estimates at 68\% CL on $b$ assuming BBH source on every specific redshift within the ET sensitivity. Middle panel : Same as left panel, but for BBHNS. Right panel: Same as middle panel, but for a mock catalogue with BNS. All events assumed here have SNR $>$ 8. The black solid line at $b = 0$ corresponds to GR.}
\label{prior_b}
\end{figure*}

In this section, we summarize how future GW detections will constrain $f(T)$ theories relaxing the condition $b = 0$ as input value on the GWs catalog simulation.

It is well known that a particular choice for the value of the parameter can bias the results in the forecast analysis. Otherwise, a natural input value for the parameter is the one predicted by the most accepted theory, GR in our case, justifying our choice $b=0$ in the preceding sections. Let us now relax this assumption by supposing that $b$ lies in the interval $b \in [-0.01, 0.01]$, which is in complete agreement with the current constraints on $f(T)$ gravity, and let us generate simulated GWs catalogs events with the same specifications as before, but with $b$ sampled in this range.

Figure \ref{aligo_prior_b} shows the estimates at 68\% CL on $b$ assuming BBH source on every specific redshift within the range $z \in [0, 1]$ from the perspective of the aLIGO sensitivity. Except for possible very low $z$ GWs events, with high SNR values (see figure \ref{SNR_aligo_BH} for better clarification), most of the events to be measured at highs $z$ are compatible with $b=0$ at significant statistical reliability, e.g, $>$ 95\% CL.

Figure \ref{prior_b} shows the constraints at 68\% CL on $b$ within the ET power spectral density noise. Once ET will be up to 10 times more sensitive in amplitude than aLIGO, it will be also able to measure GW events with higher SNR, leading to great precision on the intrinsic parameters of the theory. We can note that the estimates on $b$ from BBH (left panel), that its majority (to be at low or high $z$) can constrain b $\neq 0$ even at 99\% CL. The new borders on $b$ from BBHNS (middle panel) and BHS (right panel) for GW events at high $z$ are well compatible as $b \simeq 0$. On the other hand, GW events at very low $z$ can lead to significant non-null values on $b$.

It is important to mention that we are not taking into account possible systematic effects in our analysis to simulate the GW events at low $z$. Our estimates are taking only instrumental errors on each GW events. For instance, peculiar velocity due to the clustering of galaxies and weak lensing effects, are factors which can induce contributions on the estimates of the error bar on the free parameters of the theory at low $z$ (peculiar velocity) and high $z$ (weak lensing), respectively. Possibly taking into account these systematic effects, will be possible for checking deviations from the GR within $f(T)$ gravity only from BBH events via ET sensitivity.

\subsection{Luminosity distance from GWs}

\begin{figure*}
\includegraphics[width=3.0in,height=2.5in]{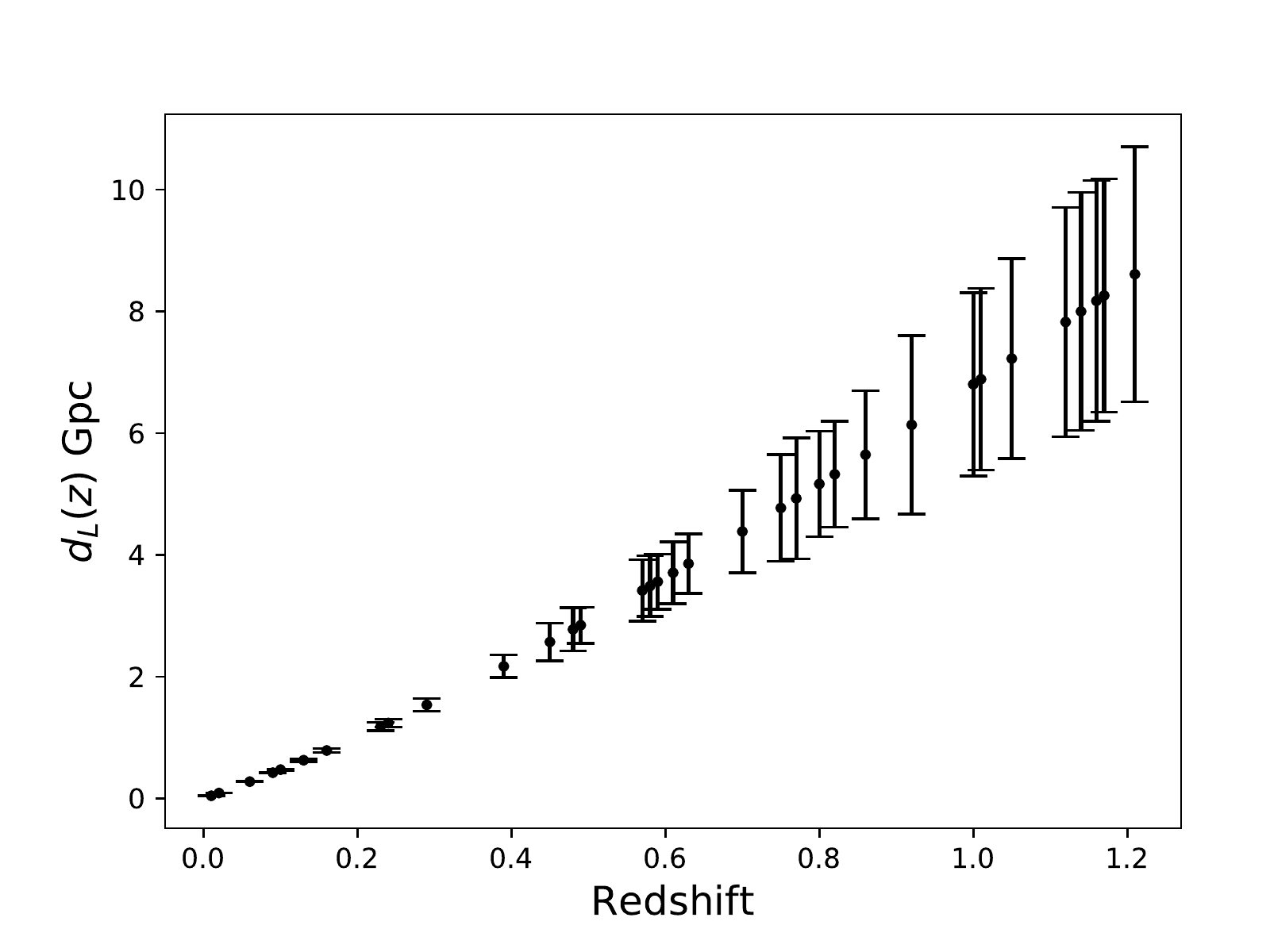} \quad \quad
\includegraphics[width=3.0in,height=2.5in]{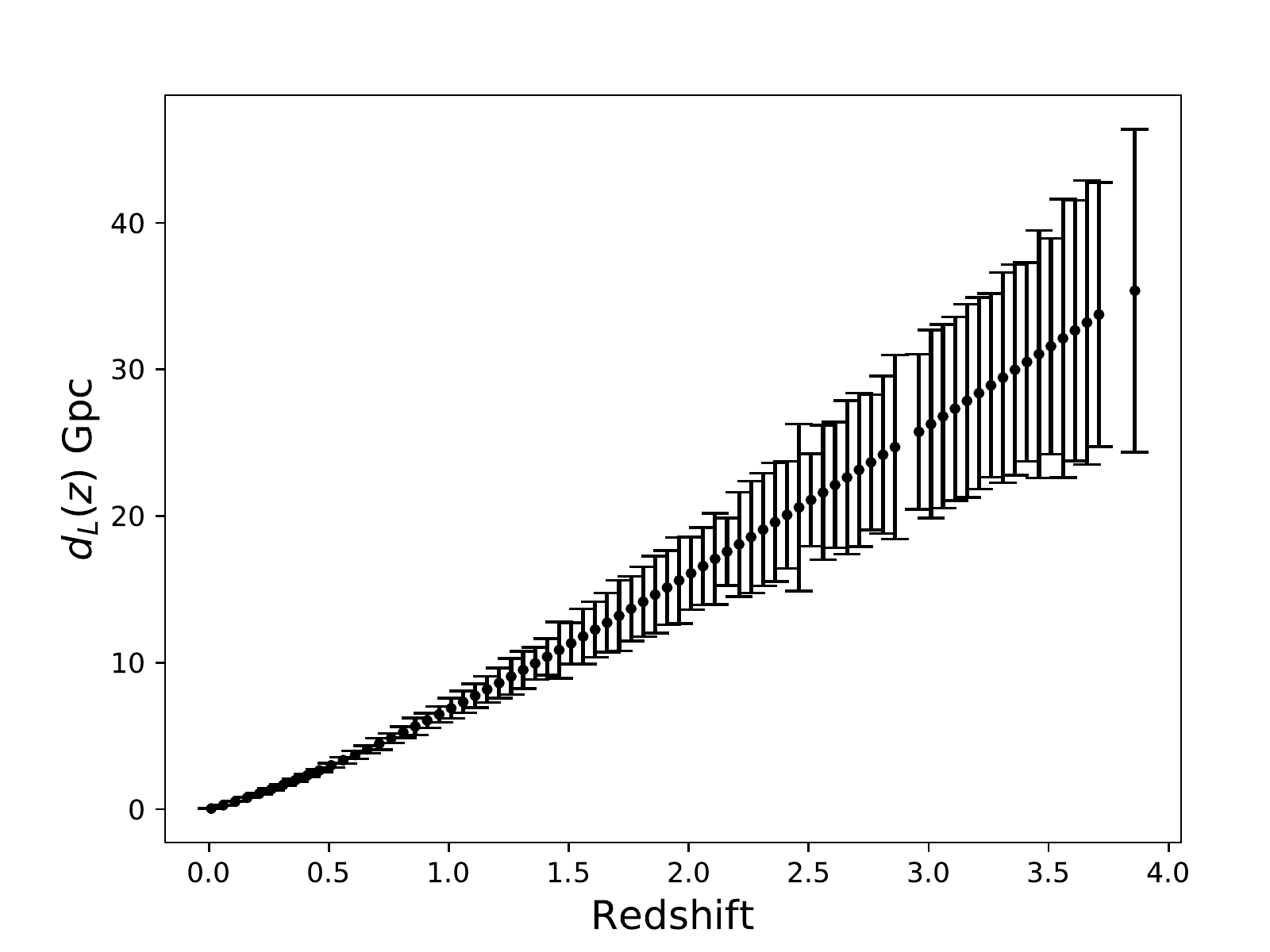}
\caption{Left panel: Catalog with observed events of luminosity distance from BNS from our fiducial $f(T)$ gravity model. Right panel: Same as in the left panel, but assumed a mock of BBHNS events.}
\label{ET_dL}
\end{figure*}

Lastly, let us discuss the GWs as a standard sirens in order to determine the luminosity distance from their detection in the context of the $f(T)$ gravity. It is important to note that the approach developed in this section is different from what was presented in the previous section regarding the estimate of the parameter $b$. This section provides a guide for readers who are interested in using standard sirens to do forecast analysis on any $f(T)$ parametric theory. Therefore, it may be of general interest of the community. In addition, this methodology can also be adapted to any GW detectors, such as DECIGO, LISA, etc.

Following a standard procedure, we now apply the Fisher matrix to estimate the error on the measurement of the luminosity distance within the $f(T)$ modified gravity context. Let us denote the luminosity distance by $d_L$ (as usual), thus it is  only to replace the notation of $d_L^{GW}$. 

Assuming that the errors on $d_L$ are uncorrelated with errors on the remaining GW parameters, we have
\begin{equation}
\sigma_{d_L}^2 = \Big( \frac{\partial \tilde{h}(f)}{\partial d_L}, \frac{\partial \tilde{h}(f)}{\partial d_L} \Big)^{-1}.
\end{equation}

Once that $\tilde{h}(f) \propto d_L^{-1}$, hence $\sigma_{d_L} =  d_L/ \rho$. Note that $\tilde{h}(f)$ for $f(T)$ gravity is completely defined by Equations (\ref{waveform} - \ref{phi}). The amount $\rho$ (the SNR) from a GW signal in inspiraling of compact binary systems within $f(T)$ gravity can be evaluated via Equation (\ref{SNR}).

However, when we estimate the practical uncertainty of the measurements of $d_L$, we should take the orbital inclination into account. 
The maximal effect of the inclination on the SNR is a factor of 2 (between $\iota =0^{\circ}$ and $\iota = 90^{\circ}$). Therefore, we add this factor to the instrumental error for a conservative estimation $\sigma_{d_L} =  2 d_L/ \rho$. Another error that we need to consider is $\sigma_{d_L}^{lens}$ due to the effect of weak lensing, and we assume $\sigma_{d_L}^{lens} = 0.05 z d_L $ as in \cite{ET03, ET04}. Thus, for ET, the total uncertainty on the luminosity distance is given by

\begin{equation}
\sigma_{d_L}^2 = \Big( \frac{2 d_L}{\rho} \Big)^2 + (0.05 z d_L)^2. 
\end{equation}

In order to generate a mock catalog using modified gravity, it is necessary to enter with non-null values for the parameter $b$. Hence, let us consider the following reasonable choice,  $b = 0.01$. Figure \ref{ET_dL} shows an example of a catalogue with observed events of luminosity distance for a BNS from our fiducial modified gravity model. 

Such catalogs are quite general, and can be used to investigate any class of models or properties within $f(T)$ gravity. For examples of the usage of mock data of luminosity distance within GR in investigations in several different contexts, see, e.g. \cite{ET05,ET06, ET07,ET08,ET09,ET10,ET11,ET12,ET13,ET14,ET15,ET16,ET17,ET18,ET19,ET20,ET21,ET22,ET23,ET24,ET25,ET26}. As already argued, our estimates on $b$ in the previous sections are directly evaluated from the GW signal/strain $\tilde{h}(f)$, because such an approach is believed to be more robust. On the other hand, a mock catalog for the luminosity distance versus $z$ also can be used to investigate any aspect of the $f(T)$ gravity which is related to such a geometric test, in the same manner it is usually done for the GR theory.

\section{Final remarks}
\label{Conclusions}

The main result of the present study is that future ground based detections of high redshift GWs from binary systems are very promissing in testing the theory of gravity. The sensitivity achieved by the ET detector or a similar third generation interferometer is enough to improve the current estimates on the free parameter within $f(T)$ gravity up to two orders of magnitude. On the other hand, in the case of detections with the aLIGO, our forecast analysis indicates that is possible
to constraint the free parameter of the theory similarly to those imposed via current cosmological probes. Thus, it may be interesting to use GW data from aLIGO in possible joint analysis in the presence of another cosmological tests in future investigations, to break possible degeneracy in cosmological parameters, once that these events begin to be detected.

We have also obtained some mock catalog for luminosity distance measurements and their estimated errors, which can also be used to probe any parametric $f(T)$ functions present in the literature, which can be of general interest to community.

Notice that all the constraints were derived by evaluating directly the induced modifications in the waveform $\tilde{h}(f)$ from the inspiraling of compact binary systems due to a change in the gravity theory. With such an approach, it is possible to have stronger constraints in the case of few detected events. To apply similar methodology as developed here can be promising to investigate some general theories of gravity \cite{nos} and to derive new statistical expectations on such models, whether by GWs signal from binary systems or from a primordial stochastic background. Also, the development of the parameterized post-Einsteinian framework for $f(T)$ gravity as a generalization and a direct comparison with the results presented here will be promising.
\\

\textbf{Note}: Interested in use any of the simulated data presented here, in particular to get any mock catalog with observed events of luminosity distance. Contact us for information.
\\

\begin{acknowledgments}
\noindent 
The authors would like to thank Massimo Tinto and Emmanuel N. Saridakis for very useful comments. RCN would like to thank the agency FAPESP for financial support under the project \# 2018/18036-5. MESA and JCNA would like to thank the Brazilian agency FAPESP for financial support under the thematic project \# 2013/26258-4. JCNA would like to thank CNPq for partial financial support under grant \# 307217/2016-7.

\end{acknowledgments}

%%%%%%%%%%%%%%%%%%%%%%%%%%%%%%%%%%%%%%%%%%%%%%%%%%%%%%%%%%%%%%%%%%%%%%%%%%%%%%%%%%%%%%%%%%%%%%%%%%%%%%%%%%%%%%%%%%%%
%%%%%%%%%%%%%%%%%%%%%%%%%%%%%%%%%%%%%%%%%%%%%%%%%%%%%%%%%%%%%%%%%%%%%%%%%%%%%%%%%%%%%%%%%%%%%%%%%%%%%%%%%%%%%%%%%%%%

\end{document}